\documentclass[aps,pra,twocolumn,groupedaddress]{revtex4-2}
\usepackage{amsmath}
\usepackage{amssymb}
\usepackage{xcolor}
\usepackage{commath}
\usepackage{graphicx}
\usepackage{hyperref}
\DeclareMathOperator{\tr}{tr}

\begin{document}

\title{Saturable Purcell filter for circuit quantum electrodynamics}

\author{Ivan Iakoupov}
\author{Kazuki Koshino}
\email[kazuki.koshino@osamember.org]{}
\affiliation{College of Liberal Arts and Sciences, Tokyo Medical and Dental University, 2-8-30 Konodai, Ichikawa, Chiba 272-0827, Japan}

\date{\today}

\begin{abstract}
We consider a typical circuit QED setup where an artificial atom encodes a 
qubit and is dispersively coupled to a measurement resonator that in turn is 
coupled to a transmission line. We show theoretically that by placing another 
artificial atom in this transmission line to act as a filter, the Purcell 
decay of the qubit into the transmission line is suppressed. When strong 
control fields are applied in the transmission line, the filter is saturated 
and effectively switched off. Such a Purcell filtering capability permits both 
the control and measurement of the qubit using the single transmission line, 
while maintaining the long coherence time of the qubit in the absence of the 
control pulses. We show that high fidelity Pauli $\sigma_x$ gates on the qubit 
can be realized using simple pulse shapes. For devices that already use one 
transmission line both for control and measurement of the qubit, our work 
provides a way to completely filter out the qubit frequency without removing 
the possibility of controlling the system. Further, combining the proposed 
filter with frequency multiplexing potentially enables both control and 
measurement of several qubits using a single Purcell-filtered transmission 
line. This will enhance the scalability of superconducting quantum processors 
by decreasing the number of the required transmission lines.
\end{abstract}

\maketitle
\section{Introduction}
Superconducting quantum processors with moderate numbers of qubits are already 
available~\cite{arute_nature19,jurcevic_qst21,gong_science21,zhu_sb22,zhang_sciadv22}. 
Error correction, where multiple physical qubits act as one logical qubit, is already 
being explored on such scalable hardware~\cite{chen_prl22,zhao_prl22,acharya_2022}. This means, however, that the 
number of the physical qubits has to increase drastically to be able to run useful 
quantum algorithms on the logical qubits.
To do this, every part of the current setups 
needs to be improved: the artificial atoms 
that encode the qubits, the room-temperature electronics that
controls them, and the interconnect.
Simplifying the interconnect is the 
focus of this article. In particular, we will show how to reduce the number of 
the microwave transmission lines. The current approach is to use frequency 
multiplexing where several qubits are measured using the same 
transmission
line~\cite{arute_nature19,jurcevic_qst21,gong_science21,zhu_sb22,zhang_sciadv22}.
Each qubit is coupled to the measurement line through a resonator 
that has a significantly different frequency.
The measurement line is filtered to suppress the qubit decay into 
it. Such decay is conventionally called ``Purcell 
decay''~\cite{reed_apl10}, and the filters are called ``Purcell filters''.

\begin{figure}[t]
\begin{center}
\includegraphics{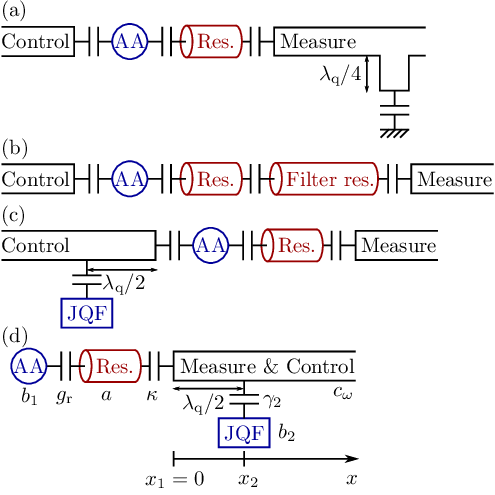}
\end{center}
\caption{(a)~Band-rejection Purcell filtering using a stub 
filter~\cite{reed_apl10}. The qubit is encoded in the coupled eigenstates 
(dressed states) of an 
artificial atom (AA) and a resonator (Res.). The 
measurement line is connected to the resonator and has an open circuit stub with length 
$\lambda_\text{q}/4$, where $\lambda_\text{q}$ is the wavelength 
corresponding to the qubit transition. Alternative designs of the 
band-rejection filters can be found in Refs.~\cite{bronn_apl15,sunada_prapplied22}.
(b)~Bandpass Purcell filtering using a filter resonator~\cite{jeffrey_prl14}. 
The frequencies of the resonators are the same or close to each other. (c)~Filtering of 
the control line using a Josephson quantum 
filter~(JQF)~\cite{koshino_prapplied20,kono_ncomms20} placed 
$\lambda_\text{q}/2$ away from the qubit. The Purcell filter in 
the measurement line is not shown. (d)~The setup that we consider---JQF as a 
saturable Purcell filter in the combined measurement and control line.
\label{fig_filters}}
\end{figure}

The Purcell filters are designed to break the trade-off between fast 
measurement and small Purcell decay. This is done by 
filtering the frequencies close to the qubit transitions, but not the 
resonator 
frequencies~\cite{reed_apl10,jeffrey_prl14,bronn_apl15,sunada_prapplied22} 
[see Figs.~\ref{fig_filters}(a) and
\ref{fig_filters}(b)]. Addition of an unsaturable Purcell 
filter to a transmission line makes 
it more challenging to control the qubits, precisely due to the fact that 
the qubit frequencies are filtered out. If they are filtered out 
completely, a separate (unfiltered) control line is 
required for each qubit~\cite{arute_nature19,jurcevic_qst21,gong_science21,zhu_sb22,zhang_sciadv22}.
However, it is possible to make a trade-off between filtering and leaving a small 
coupling to perform control using the measurement line~\cite{sunada_prapplied22}.

The qubit transition frequencies could not be 
filtered out completely before the introduction of the Josephson quantum filter 
(JQF)~\cite{koshino_prapplied20,kono_ncomms20}, which is another artificial 
atom. The JQF 
matches the qubit transition (i.e., the transition frequency of the 
two lowest energy levels the JQF is approximately 
the same as the qubit transition), is strongly coupled to the control line, and is 
placed half a wavelength apart as shown in Fig.~\ref{fig_filters}(c). When a 
strong control pulse is applied, the JQF becomes saturated and effectively 
switched off. In the absence of the control pulses, the JQF prevents the decay 
of the qubit into the control line. 
Therefore, the JQF breaks the trade-off between fast control and small decay 
rate of the qubit into the control line.

In this article, we show that the JQF can also act 
as a Purcell filter when placed in the measurement line. Because the JQF can 
be saturated, it allows resonant control pulses to be sent 
in the measurement line, making the separate control lines unnecessary. This 
results in the setup shown in Fig.~\ref{fig_filters}(d). We verify that simple 
control pulses are sufficient to implement high-fidelity gates on the qubit, and that the 
gate fidelity can be further increased using quantum optimal control. We also briefly 
comment on the combination of the JQFs with the frequency multiplexing. Such a combination would have the same low number of transmission lines as in 
Refs.~\cite{jerger_apl12,chen_apl12,rosenberg_npjqi17}, but with the Purcell 
filtering of the qubits. We expect that the Purcell filtering using a JQF will also be useful outside of the quantum 
computation context, e.g., in the hybrid systems setup of 
Ref.~\cite{lachance-quirion_science20}.

The rest of the article is organized as follows. In Sec.~\ref{sec_setup} we 
describe our theoretical model and its associated parameters. In 
Sec.~\ref{sec_decay}, we show that the Purcell decay is suppressed by the JQF. 
In Sec.~\ref{sec_measurement}, we show that the JQF does not disturb the 
measurement of the qubit. In Sec.~\ref{sec_control}, we show that the qubit can be 
controlled despite the presence of the JQF by finding the pulse shapes that 
implement a high-fidelity Pauli $\sigma_x$ gate. The article is concluded by 
Sec.~\ref{sec_conclusion}.

\section{Setup}\label{sec_setup}
We assume transmons, modeled as anharmonic oscillators, for the artificial 
atoms. Using the two-level atoms gives similar 
results~\cite{koshino_prapplied20,kono_ncomms20}. The considered setup is 
shown in Fig.~\ref{fig_filters}(d). One transmon is 
coupled to a transmission line through a resonator. The subsystem consisting 
of this transmon and resonator has index~$1$ in the equations below and 
encodes a qubit in its coupled eigenstates (dressed states).
The subsystem~$2$ is a transmon that is coupled to the same 
line directly and is used as a JQF. The JQF is placed $\lambda_\text{q}/2$ 
from the resonator, where $\lambda_\text{q}$ is the 
wavelength that corresponds to the qubit transition 
frequency. The Hamiltonian of the 
setup can be written $H=H_\text{s}+H_\text{f}+H_\text{i}$. The parts 
$H_\text{s}$, $H_\text{f}$, and $H_\text{i}$ correspond to the system, 
transmission line field, and the interaction, respectively.

The system part is $H_\text{s}=\sum_{m=1}^2 H_{\text{s},m}$, where the 
subsystem $1$ part is
\begin{align}\label{H_sm_aa_res}
\begin{aligned}
&H_{\text{s},1}=
\hbar\omega_{\text{t},1} b_1^\dagger b_1
+\hbar\frac{\alpha_1}{2} (b_1^\dagger)^2 b_1^2\\
&+\hbar\omega_\text{r} a^\dagger a
+\hbar g_\text{r}
(b_1^\dagger a + a^\dagger b_1),
\end{aligned}
\end{align}
and the subsystem $2$ part is
\begin{align}\label{H_sm_jqf}
&H_{\text{s},2}=
\hbar\omega_{\text{t},2} b_2^\dagger b_2
+\hbar\frac{\alpha_2}{2} (b_2^\dagger)^2 b_2^2.
\end{align}
The resonator has the corresponding annihilation operator~$a$ and the frequency~$\omega_\text{r}$. There are $2$ transmons 
with the corresponding 
annihilation operators~$b_m$, 
transition frequencies~$\omega_{\text{t},m}$ of the lowest two energy levels, 
and the anharmonicity parameters~$\alpha_m$. Only the transmon with the index 
$m=1$ is coupled to the resonator with 
the coupling strength~$g_\text{r}$. 

Multiple resonator modes could have a significant contribution to the Purcell 
decay~\cite{houck_prl08,lachance-quirion_science20}. This occurs when the 
detunings of several modes of the resonator from the qubit frequency have 
similar magnitudes. It is possible to model a multimode resonator as several 
single-mode ones~\cite{malekakhlagh_pra16,malekakhlagh_prl17,gely_pra17}, but 
for simplicity, we only consider the parameter regime where one of the 
resonator modes is dominant. We use the parameters close to 
Ref.~\cite{kono_ncomms20}, where a coplanar waveguide resonator with the 
fundamental frequency $\omega_\text{r}/(2\pi)=10\text{ GHz}$ was coupled to 
the transmon with the transition 
frequency~$\omega_{\text{t},1}/(2\pi)=8\text{ GHz}$. The next mode of the resonator has the frequency 
$2\omega_\text{r}/(2\pi)=20\text{ GHz}$, and hence does not contribute much to 
the Purcell decay.

The existence of the higher modes does not significantly change the 
required JQF parameters. The JQF needs to have the frequency close to 
the qubit frequency, not the resonator modes. The resonator modes shift 
the qubit frequency due to the coupling, but the shift due 
to the fundamental mode is a few MHz for the considered parameters, and the 
shift due to the higher modes is
even smaller. In practice, the frequency of the JQF may need to be tuned post 
fabrication anyway, either using the bias flux~\cite{kono_ncomms20} or the 
laser annealing~\cite{zhang_sciadv22}. Hence, even the predictions of a 
single-mode theory should be sufficient for this parameter regime.

The transmission line field part is
\begin{align}
&H_\text{f}=\hbar\int_0^\infty \omega c_\omega^\dagger c_\omega \dif \omega.
\end{align}
The annihilation operators~$c_\omega$ correspond to
the modes $\cos(k_\omega x)$ with positive wave vectors~$k_\omega$, but use the angular 
frequencies~$\omega$ as the integration variable. The dispersion relation is 
$\omega=k_\omega v_\text{g}$, with $v_\text{g}$ being the 
speed of light (group velocity) in the transmission line.

The interaction part is
\begin{align}\label{H_i}
&\begin{aligned}
H_\text{i}=-\hbar\sum_{m=1}^{2} \int_0^\infty
g_m(\omega)
(c_\omega-c_\omega^\dagger)(\mathcal{O}_m  - \mathcal{O}_m^\dagger)\dif \omega,
\end{aligned}
\end{align}
where $g_m(\omega)=G_m\sqrt{\omega}\cos(k_\omega x_m)$, $G_m=\sqrt{\Gamma_m/(2\pi\omega_m)}$,
\begin{align}\label{Gamma_m_def}
&\Gamma_m=\left\{\begin{aligned}
&\kappa\text{ for } m=1,\\
&\gamma_2\text{ for } m=2,
\end{aligned}\right.\\\label{omega_m_def}
&\omega_m=\left\{\begin{aligned}
&\omega_{\text{r}}\text{ for } m=1,\\
&\omega_{\text{t},2}\text{ for } m=2,
\end{aligned}\right.\\\label{O_m_def}
&\mathcal{O}_m=\left\{\begin{aligned}
&a\text{ for } m=1,\\
&b_2\text{ for } m=2.
\end{aligned}\right.
\end{align}

The interaction Hamiltonian~\eqref{H_i} gives rise to the 
decay rates $\kappa$ (resonator) and $\gamma_2$ (JQF). We use the coupling of 
the form
$g_m(\omega)=G_m\sqrt{\omega}\cos(k_\omega x_m)$, which is obtained by ignoring the so-called 
$A^2$~term. A more careful derivation~\cite{bamba_pra14,malekakhlagh_pra16,malekakhlagh_prl17,gely_pra17} results in 
$g_m(\omega)=G_m(\sqrt{\omega}/\sqrt{1+\mathcal{A}\omega^2})\cos(k_\omega x_m)$, i.e., has a 
cutoff for the higher $\omega$ controlled by the parameter $\mathcal{A}>0$. The form of $g_m$ with the cutoff results in the 
renormalization of the decay rates and an additional
collective frequency shift as detailed in App.~\ref{App:cutoff_in_coupling}. 
The theoretical model without a cutoff, i.e., with  
$g_m(\omega)=G_m\sqrt{\omega}\cos(k_\omega x_m)$, was 
found to be in good agreement with the experiment when the JQF was placed in 
the dedicated control line~\cite{kono_ncomms20}, suggesting that the influence 
of the 
additional frequency shift is small. To estimate its effect 
theoretically, the numerical value of $\mathcal{A}$ is needed, and it does not 
seem to be available in the literature. Due to these considerations, we 
proceed with $g_m(\omega)=G_m\sqrt{\omega}\cos(k_\omega x_m)$, but the calculations could be 
easily adjusted for a non-zero $\mathcal{A}$.

The constants $G_m$ could also be related to the circuit parameters, but we 
write them in terms of the decay rates $\Gamma_m$, which can be 
measured experimentally~\cite{kono_ncomms20}. If the $A^2$ terms are included, 
these constants could be modified to $G_m=\sqrt{\Gamma_m(1+\mathcal{A}\omega_{m}^2)/(2\pi\omega_m)}$ to account 
for the renormalization.
Another note is that the rotating wave approximation is not applied immediately in $H_\text{i}$. 
It will be applied after the effective Heisenberg equations of motion 
for the subsystems are obtained to ensure that all the 
terms are present~\cite{ott_pra13}. More details about this can be found in 
App.~\ref{App:master_equation} and App.~\ref{App:cutoff_in_coupling} where the 
master equation is derived starting from the Hamiltonian above and following 
Refs.~\cite{koshino_prapplied20,kono_ncomms20,ott_pra13,lehmberg_pra70}. Here, 
we only give the outline of this derivation.

In general, the interaction of matter with the electromagnetic fields results 
in non-Markovian equations of motion~\cite{wodkiewicz_ap76,deVega_rmp17} 
caused by the fact that 
it takes a finite time for the photons to propagate between the atoms. For 
$g_m(\omega)=G_m\sqrt{\omega}\cos(k_\omega x_m)$, the 
effective equations of motion where the field degrees of freedom are traced 
out, take the form of the delay differential equations. However, the delay 
differential equations are difficult to solve in the general case, 
and hence some kind of approximation is usually needed. We adopt the 
approximation that converts the time delays into the propagation phase 
factors~\cite{koshino_prapplied20,kono_ncomms20,ott_pra13,lehmberg_pra70}.
For 
$g_m(\omega)=G_m(\sqrt{\omega}/\sqrt{1+\mathcal{A}\omega^2})\cos(k_\omega x_m)$,
the equations of motion are of a more general form with a memory kernel given 
by Eq.~\eqref{delta_prime_integral_cutoff_kernel} in 
App.~\ref{App:master_equation}. This case can also be approximated by a 
Markovian master equation as explained in App.~\ref{App:cutoff_in_coupling}.

Before we explain the approximation involved in replacing the non-Markovian 
equations of motion with the Markovian ones, we first note that we
diagonalize the subsystem Hamiltonians $H_{\text{s},m}$. For every pair of eigenstates $|j_m\rangle$, $|j'_m\rangle$ of 
$H_{\text{s},m}$, we define the operators 
$\sigma_{m,jj'}=|j_m\rangle\langle j'_m|$ and the 
matrix elements $C_{m,jj'}=\langle j_m|\mathcal{O}_m|j'_m\rangle$. We order 
the eigenstates such that the number of excitations increases or is constant 
with increasing $j$. Since $\mathcal{O}_m$ 
is an annihilation operator, the 
rotating wave approximation ensures that $C_{m,jj'}\neq 0$ only for $j<j'$. We can
write 
\begin{gather}\label{H_sm_diag_basis}
H_{\text{s},m}=\hbar\sum_{j}\omega_{m,j}\sigma_{m,jj},
\end{gather}
where $\omega_{m,j}$ are the eigenfrequencies.

Assuming that the Hamiltonian is dominated by the 
system parts~\eqref{H_sm_diag_basis}, the approximation of the time-delayed terms can be written
\begin{gather}\label{free_evolution_appr}
\sigma_{m,jj'}(t-t_x)\approx\sigma_{m,jj'}(t)e^{i(\omega_{m,j'}-\omega_{m,j})t_x}, 
\end{gather}
making the Heisenberg 
equations of motion for the attached subsystems local in time. When a classical drive 
with frequency $\omega_\text{d}$ is present, we make the approximation 
\begin{gather}\label{driven_evolution_appr}
\sigma_{m,jj'}(t-t_x)\approx\sigma_{m,jj'}(t)e^{i\omega_\text{d}t_x} 
\end{gather}
instead. Physically, this means that the driven subsystems oscillate with the drive frequency 
rather than their eigenfrequencies. 

The derivation assumes a coherent state with the carrier frequency 
$\omega_\text{d}$ as the input in the transmission line, and hence an 
additional drive Hamiltonian
\begin{gather}\label{drive_Hamiltonian}
H_\text{d}
=\hbar\sum_{m=1}^2
\left(\Omega_m e^{-i\omega_\text{d} t} \mathcal{O}_m^\dagger
+ (\Omega_m)^* e^{i\omega_\text{d} t} \mathcal{O}_m\right)
\end{gather}
emerges with the Rabi frequencies
\begin{gather}\label{Omega_m_def}
\Omega_m
=\sqrt{\frac{\omega_\text{d}}{\omega_m}\Gamma_m\dot{n}}\cos(k_{\omega_\text{d}} x_m)e^{i\phi},
\end{gather}
which may be time-dependent due to the changing photon
flux~$\dot{n}$ and phase $\phi$.

The rotating 
frame is defined with respect to the Hamiltonian
\begin{gather}\label{H_0_def}
H_0=\hbar\sum_{m=1}^2\sum_{j} \omega_{0,m,j} \sigma_{m,jj},
\end{gather}
where the frequencies $\omega_{0,m,j}$ are chosen such that the factors $e^{\pm i\omega_\text{d} t}$ in Eq.~\eqref{drive_Hamiltonian}
are canceled, i.e., 
$\frac{i}{\hbar}[H_0,\mathcal{O}_m]=-i\omega_\text{d}\mathcal{O}_m$. If there 
is no drive, any fixed frequency can be used instead of $\omega_\text{d}$. The 
Hamiltonian in the rotating frame is
\begin{gather}\label{Hamiltonian_rotating_frame}
\begin{aligned}
&\tilde{H}=\hbar\sum_{m=1}^2\sum_{j} (\omega_{m,j}-\omega_{0,m,j}) \sigma_{m,jj}\\
&+\text{Re}[\Omega]\tilde{H}_\text{d,Re}+\text{Im}[\Omega]\tilde{H}_\text{d,Im},
\end{aligned}
\end{gather}
where we have picked $\Omega=\Omega_1$ as the reference Rabi frequency. 
Defining
\begin{gather}
\tilde{\Omega}_m=\frac{\text{Re}[\Omega_m]}{\text{Re}[\Omega]}=\frac{\text{Im}[\Omega_m]}{\text{Im}[\Omega]}
=\sqrt{\frac{\omega_{m_0}}{\omega_m}\frac{\Gamma_m}{\Gamma_{m_0}}}\frac{\cos(k_{\omega_\text{d}} x_m)}{\cos(k_{\omega_\text{d}} x_{m_0})},
\end{gather}
we can write
\begin{subequations}
\begin{align}
&\tilde{H}_\text{d,Re}=\hbar\sum_{m=1}^2
\tilde{\Omega}_m
\left(\mathcal{O}_m^\dagger + \mathcal{O}_m\right),\displaybreak[0]\\
&\tilde{H}_\text{d,Im}=i\hbar\sum_{m=1}^2
\tilde{\Omega}_m
\left(\mathcal{O}_m^\dagger - \mathcal{O}_m\right).
\end{align}
\label{H_d_Re_Im_def}
\end{subequations}
The ratios $\tilde{\Omega}_m$ do not depend on the photon flux 
$\dot{n}$ or phase $\phi$, and hence $\tilde{H}_\text{d,Re}$ and $\tilde{H}_\text{d,Im}$ are independent of time. The time dependence of $\tilde{H}$ is 
contained in the factors $\text{Re}[\Omega]$ and $\text{Im}[\Omega]$. 

The master equation can be written
\begin{gather}\label{master_equation}
\begin{aligned}
&\dot{\tilde{\rho}}_\text{s}
=\mathcal{L}(\tilde{\rho}_\text{s})
=-\frac{i}{\hbar}[\tilde{H},\tilde{\rho}_\text{s}]\\
&+\frac{1}{2}\sum_{m,n=1}^2\left(\mathcal{O}_{mn}\tilde{\rho}_\text{s}\mathcal{O}_m^\dagger-\mathcal{O}_m^\dagger\mathcal{O}_{mn}\tilde{\rho}_\text{s}\right)\\
&+\frac{1}{2}\sum_{m,n=1}^2\left(\mathcal{O}_n\tilde{\rho}_\text{s}\mathcal{O}_{nm}^\dagger-\tilde{\rho}_\text{s}\mathcal{O}_{nm}^\dagger\mathcal{O}_n\right),
\end{aligned}
\end{gather}
where $\mathcal{O}_{mn}=\sum_{j,j'}\xi_{mn,j'j}C_{n,jj'}\sigma_{n,jj'}$,
\begin{gather}\label{xi_def}
\begin{aligned}
&\xi_{mn,j'j}=\frac{\sqrt{\Gamma_m \Gamma_n}}{2} \frac{\omega_{n,j'j}}{\sqrt{\omega_m\omega_n}}\\
&\times\left(e^{ik_{n,j'j} |x_m-x_n|}+e^{ik_{n,j'j} |x_m+x_n|}\right),
\end{aligned}
\end{gather}
$\omega_{n,j'j}=\omega_{n,j'}-\omega_{n,j}$ are the transition frequencies 
between the eigenstates $j$ and $j'$, 
$k_{n,j'j}=k_{\omega_{n,j'j}}$ are the corresponding wavevectors, 
and $\tilde{\rho}_\text{s}=e^{iH_0t/\hbar}\tr_\text{f}[\rho]e^{-iH_0t/\hbar}$ is 
the density matrix with the transmission line field degrees of freedom traced out, in 
the rotating frame with respect to the Hamiltonian~\eqref{H_0_def}. If a 
classical drive is present, we set $k_{n,j'j}=k_{\omega_\text{d}}$ in 
Eq.~\eqref{xi_def}, while keeping the factor
$\omega_{n,j'j}/\sqrt{\omega_m\omega_n}$ unchanged. For
$g_m(\omega)=G_m(\sqrt{\omega}/\sqrt{1+\mathcal{A}\omega^2})\cos(k_\omega x_m)$ 
with $\mathcal{A}>0$, $\xi_{mn,j'j}$ is given by Eq.~\eqref{xi_def_A} derived 
in App.~\ref{App:cutoff_in_coupling} instead of Eq.~\eqref{xi_def}.

The Schrieffer-Wolff transformation on the 
Hamiltonian~\eqref{H_sm_aa_res} results in the dispersive shifts for every 
transmon energy level~\cite{gambetta_notes13}. We define
\begin{gather}\label{chi_definition}
\chi=\frac{g_\text{r}^2}{2(\omega_\text{r}-\omega_{\text{t},1})}\left(1-\frac{\omega_\text{r}-\omega_{\text{t},1}+\alpha_1}{\omega_\text{r}-\omega_{\text{t},1}-\alpha_1}\right)
\end{gather}
in terms of the difference of the dispersive shifts for the lowest two levels. 
For $\alpha_1\rightarrow\infty$, this reduces to the two-level system 
shift $\chi=g_\text{r}^2/(\omega_\text{r}-\omega_{\text{t},1})$.

For the calculations below, the parameters are chosen close to the ones in 
Ref.~\cite{kono_ncomms20}. We set the frequency of the resonator 
$\omega_\text{r}/(2\pi)=10\text{ GHz}$, resonator decay rate 
$\kappa/(2\pi)=2\text{ MHz}$, transition frequencies of the lowest two 
transmon energy levels
$\omega_{\text{t},1}/(2\pi)=8.000\text{ GHz}$ and
$\omega_{\text{t},2}/(2\pi)=7.994\text{ GHz}$ (shifted to match the qubit 
transition frequency $\omega_{1,10}/(2\pi)$),
anharmonicities $\alpha_1/(2\pi)=\alpha_2/(2\pi)=-400\text{ MHz}$, JQF decay 
rate $\gamma_2/(2\pi)=100\text{ MHz}$, and the dispersive shift
$\chi/(2\pi)=1\text{ MHz}$. From
Eq.~\eqref{chi_definition}, the coupling between the transmon and the 
resonator is 
$g_\text{r}/(2\pi)=109.544\text{ MHz}$. The resonator is placed at the origin, 
$x_1=0$, and the JQF is placed half a wavelength from the resonator, $k_{\omega_{1,10}}x_2=\pi$. 
We 
also choose $\Omega=\Omega_1$ as the reference Rabi frequency, and hence 
$\tilde{\Omega}_1=1$ and
$\tilde{\Omega}_2=\sqrt{\omega_\text{r}\gamma_2/(\omega_{\text{t},2}\kappa)}
\cos(k_{\omega_\text{d}}x_2)$.

As explained above, the 
Hamiltonian~\eqref{H_sm_aa_res} for the subsystem $1$ is written in the diagonal form~\eqref{H_sm_diag_basis} prior to the 
derivation of the master equation~\eqref{master_equation} (the 
Hamiltonian~\eqref{H_sm_jqf} for the subsystem $2$ is already diagonal). Truncated to at most one 
excitation either in the transmon or the resonator, we can write the 
Hamiltonian~\eqref{H_sm_aa_res} as the matrix
\begin{gather}
\begin{pmatrix}\label{H_sm_aa_res_single_excitation}
0 & 0 & 0 \\
0 & \omega_\text{r} & g_\text{r} \\
0 &g_\text{r} & \omega_{\text{t},1}
\end{pmatrix},
\end{gather}
where the zero row and column were added explicitly for the zero-excitation 
state $|0_1\rangle$. This is also one of the eigenstates of the matrix. The 
other two eigenstates have a single excitation and can be written
\begin{subequations}
\begin{align}
\label{state_1_1_def}
&|1_1\rangle=\sin(\theta)a^\dagger|0_1\rangle-\cos(\theta)b_1^\dagger|0_1\rangle,\displaybreak[0]\\
\label{state_2_1_def}
&|2_1\rangle=\cos(\theta)a^\dagger|0_1\rangle+\sin(\theta)b_1^\dagger|0_1\rangle,
\end{align}
\label{H_sm_aa_res_single_excitation_eigvecs}
\end{subequations}
where
$\theta=\frac{1}{2}\arg[(\omega_\text{r}-\omega_{\text{t},1})/2+ig_\text{r}]$, 
and $\arg$ is the argument of a complex number. Below, we use the 
computational basis states $|0\rangle=|0_1\rangle$ and $|1\rangle=|1_1\rangle$. 
The state $|2_1\rangle$ is the rapidly-decaying eigenstate with most of the 
excitation in the resonator, and is outside of the computational basis. The 
corresponding eigenfrequencies are $\omega_{1,0}=0$,
\begin{subequations}
\begin{align}\label{omega_1_1}
&\omega_{1,1}=\frac{\omega_\text{r}+\omega_{\text{t},1}}{2}
-\sqrt{\left(\frac{\omega_\text{r}-\omega_{\text{t},1}}{2}\right)^2+g_\text{r}^2},\\
\label{omega_1_2}
&\omega_{1,2}=\frac{\omega_\text{r}+\omega_{\text{t},1}}{2}
+\sqrt{\left(\frac{\omega_\text{r}-\omega_{\text{t},1}}{2}\right)^2+g_\text{r}^2}.
\end{align}
\label{H_sm_aa_res_single_excitation_eigvals}
\end{subequations}
With the chosen parameters above, we have 
$\omega_{1,1}/(2\pi)=7.994\text{ GHz}$ and 
$\omega_{1,2}/(2\pi)=10.006\text{ GHz}$.

The single-excitation states are sufficient to describe the decay in 
Sec.~\ref{sec_decay} below. The measurement [Sec.~\ref{sec_measurement}] and 
control [Sec.~\ref{sec_control}] involves sending 
microwave fields through the transmission line and hence can
excite the higher eigenstates. In the general case, we truncate the transmons and the resonator at a certain maximal number of excitations and then perform 
the numerical diagonalization of the resulting Hamiltonian matrices. This 
produces an eigenfrequency~$\omega_{m,j}$ for each 
eigenstate~$|j_m\rangle$. Physically, only the transition frequencies 
$\omega_{m,j'j}=\omega_{m,j'}-\omega_{m,j}$ are relevant. However,
the choice $\omega_{1,0}=0$ as the frequency of the zero-excitation 
eigenstate~$|0_1\rangle$ results in the identities $\omega_{1,10}=\omega_{1,1}$ and 
$\omega_{1,20}=\omega_{1,2}$. Hence, for the single-excitation states, the 
distinction between the absolute eigenfrequencies and the transition frequencies 
does not exist. This distinction becomes important for the higher-excitation 
states. 

\begin{figure}[t]
\begin{center}
\includegraphics{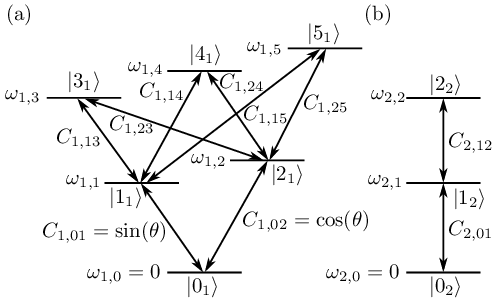}
\end{center}
\caption{(a) The level diagram of the coupled transmon and resonator subsystem, 
showing the eigenstates (dressed states) $|j_1\rangle$ up to two excitations with the corresponding 
eigenfrequencies $\omega_{1,j}$. The qubit is encoded in the states 
$|0_1\rangle$ and $|1_1\rangle$. For
$\omega_{1,1}$ and $\omega_{1,2}$, there are analytical 
expressions [cf. Eqs.~\eqref{H_sm_aa_res_single_excitation_eigvals}]. The non-zero 
matrix elements 
$C_{1,jj'}=\langle j_1|\mathcal{O}_1|j'_1\rangle=\langle j_1|a|j'_1\rangle$ 
are shown as lines with arrows between the eigenstates. For $C_{1,01}$ and 
$C_{1,02}$, there are analytical expressions using 
Eqs.~\eqref{H_sm_aa_res_single_excitation_eigvecs}. (b) The
level diagram of the transmon JQF subsystem, showing the 
eigenstates~$|j_2\rangle$ up 
to two excitations. Since the Hamiltonian~\eqref{H_sm_jqf} is already 
diagonal, the eigenfrequencies are linear combinations of its parameters:
$\omega_{2,1}=\omega_{\text{t},2}$, 
$\omega_{2,2}=2\omega_{\text{t},2}+\alpha_2$. The shown non-zero matrix elements
$C_{2,jj'}=\langle j_2|\mathcal{O}_2|j'_2\rangle=\langle j_2|b_2|j'_2\rangle$ 
are $C_{2,01}=1$ and $C_{2,12}=\sqrt{2}$. 
\label{fig_eigenstate_level_diagram}}
\end{figure}

E.g., the JQF transmon is described by an anharmonic ladder of eigenstates, as 
shown in Fig.~\ref{fig_eigenstate_level_diagram}(b). The states up 
to two excitations are $|0_2\rangle$, $|1_2\rangle$, and $|2_2\rangle$. The corresponding absolute 
eigenfrequencies are $\omega_{2,0}=0$, $\omega_{2,1}/(2\pi)=7.994\text{ GHz}$, 
and $\omega_{2,2}/(2\pi)=15.588\text{ GHz}$. The transition frequencies are 
$\omega_{2,10}/(2\pi)=7.994\text{ GHz}$ and
$\omega_{2,21}/(2\pi)=7.594\text{ GHz}$. These are the only transitions 
that have the non-zero matrix elements 
$C_{2,jj'}=\langle j_2|\mathcal{O}_2|j'_2\rangle=\langle j_2|b_2|j'_2\rangle$ 
in the two-excitation subspace. When diagonalizing the transmon 
and resonator subsystem, the non-zero matrix elements 
$C_{1,jj'}=\langle j_1|\mathcal{O}_1|j'_1\rangle=\langle j_1|a|j'_1\rangle$ have a more complicated 
structure, as shown in Fig.~\ref{fig_eigenstate_level_diagram}(a). Hence more 
transition frequencies $\omega_{m,j'j}$ are relevant to the dynamics.

\section{Suppressed Purcell decay}\label{sec_decay}

First, we verify that the decay of a qubit is reduced by adding a JQF. The computational basis state~$|0\rangle=|0_1\rangle$ is the 
zero-excitation state and hence does not 
decay. The decay of the state~$|1\rangle=|1_1\rangle$ [cf. Eq.~\eqref{state_1_1_def}] is
suppressed by the large detuning of the resonator even without a JQF. From 
the master equation~\eqref{master_equation}, the Purcell decay rate is
\begin{gather}\label{kappa_Purcell}
\kappa_\text{Purcell}=|C_{1,01}|^2\xi_{11,10}
=\sin^2(\theta)\kappa\frac{\omega_{1,10}}{\omega_\text{r}}, 
\end{gather}
where $\omega_{1,10}=\omega_{1,1}-\omega_{1,0}$, $\omega_{1,1}$ 
is given by Eq.~\eqref{omega_1_1}, and $\omega_{1,0}=0$.
Since the detuning $\omega_\text{r}-\omega_{\text{t},1}$ is large, we can 
approximate
$\sin^2(\theta)\approx (g_\text{r}/(\omega_\text{r}-\omega_{\text{t},1}))^2$, 
resulting in
$\kappa_\text{Purcell}\approx(g_\text{r}/(\omega_\text{r}-\omega_{\text{t},1}))^2
\kappa\omega_{1,10}/\omega_\text{r}$. The only difference from the
usual formula for the Purcell decay rate is an additional factor 
$\omega_{1,10}/\omega_\text{r}$. This factor comes from the definition of the frequency-dependent coupling 
$g_1(\omega)=\sqrt{\kappa/(2\pi\omega_\text{r})}\sqrt{\omega}\cos(k_\omega x_1)$
that results in the decay rate $\kappa$ of the resonator with the 
frequency $\omega_\text{r}$ that is not coupled to the transmon and placed such that 
$\cos(k_{\omega_\text{r}} x_1)=1$. The qubit transition frequency 
$\omega_{1,10}$ is different, and hence the
decay rate that is proportional to $g_1^2(\omega_{1,10})$ gets this additional 
factor $\omega_{1,10}/\omega_\text{r}$. For our parameters,
$\omega_{1,10}/\omega_\text{r}= 0.7994$, and 
$\kappa_\text{Purcell}/(2\pi)=4.8\text{ kHz}$.

\begin{figure}[t]
\begin{center}
\includegraphics{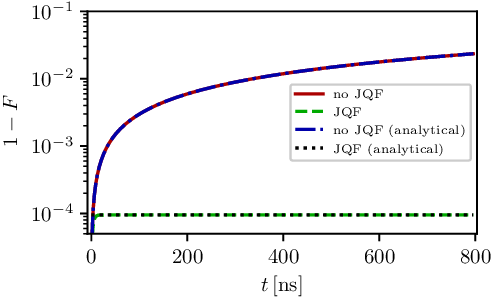}
\end{center}
\caption{The error probability $1-F$ where $F$ is given by Eq.~\eqref{F_def} as a function of time for the qubit 
initialized in the state 
$|1\rangle=|1_1\rangle$ [cf. Eq.~\eqref{state_1_1_def}], and the JQF 
initialized in the ground state $|0_2\rangle$. 
The dash-dotted blue curve plots $1-e^{-\kappa_\text{Purcell}t}$ with 
$\kappa_\text{Purcell}$ given by Eq.~\eqref{kappa_Purcell}. The horizontal dotted 
black line plots $1-F_\text{dark}$ given by Eq.~\eqref{F_dark}. It overlaps the 
dashed green curve for large times.
\label{fig_decay}}
\end{figure}

A Purcell filter needs to suppress the decay rate below $\kappa_\text{Purcell}$. 
We show in Fig.~\ref{fig_decay} that addition of a JQF accomplishes this. We 
initialize the qubit in the state~$|1\rangle=|1_1\rangle$, and the JQF (if 
present) is initialized in its ground state $|0_2\rangle$. The error 
probability $1-F$ is plotted, where the fidelity $F$ is
\begin{gather}\label{F_def}
F=\tr_\text{s}[(\sigma_{1,11}\otimes I_2)\tilde{\rho}_\text{s}(t)],
\end{gather}
and $\tr_\text{s}$ is the trace over the system degrees of freedom (the 
trace over the transmission line degrees of freedom 
$\tr_\text{f}$ has already been performed during the derivation of the 
master equation in App.~\ref{App:master_equation}).
We have also written the tensor product with the identity operator $I_2$ on the subsystem $2$ 
(JQF) in Eq.~\eqref{F_def} explicitly, so that it is more easily seen that
$F=\langle 1|\tr_2(\tilde{\rho}_\text{s})|1\rangle$,
where $\tr_2$ is the trace over the subsystem $2$.

The solid red curve in Fig.~\ref{fig_decay} is the numerically calculated 
error probability $1-F$ without a JQF, and 
the overlapping dash-dotted blue curve plots $1-e^{-\kappa_\text{Purcell}t}$.
We see that the behavior of the dashed green curve that 
shows the case with a JQF is qualitatively the same as when the JQF was placed in the dedicated control 
line~\cite{koshino_prapplied20,kono_ncomms20}. The qubit and 
the JQF have a bright state that decays rapidly and a dark 
state that does not decay. The state $|1_1\rangle\otimes|0_2\rangle$ (the qubit 
is in state~$|1\rangle$, and the JQF is in the ground state), has both 
bright and dark parts. Because the JQF decay rate $\gamma_2$ is significantly 
larger than $\kappa_\text{Purcell}$, this state is mostly dark. After the rapid decay 
of the small bright part, the decay rate 
vanishes. The fidelity after the bright part has decayed is 
\begin{gather}\label{F_dark}
F_\text{dark}=\left(\frac{\gamma_2}{\kappa_\text{Purcell}+\gamma_2}\right)^2,
\end{gather}
and $1-F_\text{dark}$ is shown by the horizontal dotted black line in 
Fig.~\ref{fig_decay}.

\begin{figure}[t]
\begin{center}
\includegraphics{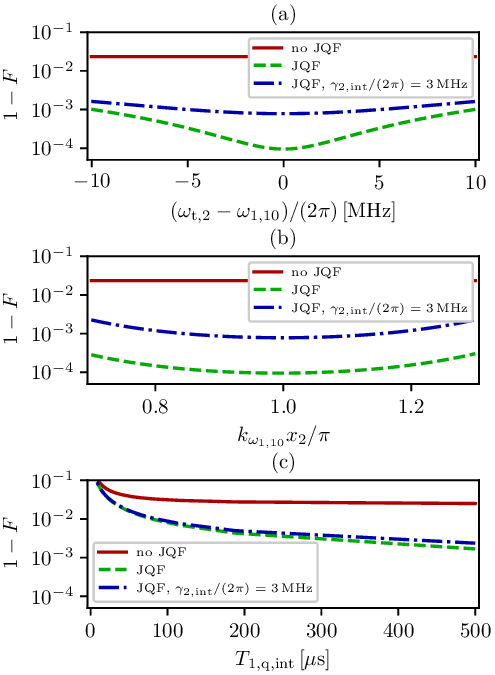}
\end{center}
\caption{The error probability $1-F$ where $F$ is given by Eq.~\eqref{F_def} as a function 
of: (a) JQF transition frequency $\omega_{\text{t},2}$, (b) JQF position $x_2$, 
(c) internal $T_1$ time of qubit ($T_{1,\text{q,int}}=1/\gamma_{\text{q,int}}$, 
where $\gamma_{\text{q,int}}$ is discussed in the text). In all the subplots, 
the qubit was initialized in the state $|1\rangle$ and evolved for the same 
time $t_\text{f}=10/\kappa\approx 800\text{ }\text{ns}$ as in 
Fig.~\ref{fig_decay}, plotting the final value. For the solid red and dashed 
green curves, the parameters are the same as in Fig.~\ref{fig_decay}, except 
for~$\omega_{\text{t},2}$ in (a), $x_2$ in (b), and $T_{1,\text{q,int}}$ in (c). The dash-dotted blue curves  
show the influence of the internal decay of the JQF with the rate 
$\gamma_{2,\text{int}}/(2\pi) = 3\text{ MHz}$~\cite{kono_ncomms20}.
\label{fig_decay_imperfections}}
\end{figure}

In practice, the decay 
rate is not expected to be zero due to the imperfections in the qubit and the JQF but still be reduced compared to the case 
without the JQF. We show the influence of some of the possible imperfections 
in Fig.~\ref{fig_decay_imperfections}. If the JQF transition 
frequency $\omega_{\text{t},2}$ does not match the qubit transition frequency $\omega_{1,10}$, the decay 
rate is not canceled completely, as 
shown in Fig.~\ref{fig_decay_imperfections}(a). The cancellation is also imperfect if 
the JQF is not placed at $x_2=\pi/k_{\omega_{1,10}}$, as shown 
in Fig.~\ref{fig_decay_imperfections}(b). The influence of the internal decay of the 
JQF with the rate $\gamma_{2,\text{int}}/(2\pi) = 3\text{ MHz}$~\cite{kono_ncomms20} is shown as the dash-dotted blue curves in Fig.~\ref{fig_decay_imperfections},
calculated by phenomenologically adding a decay term 
$\gamma_{2,\text{int}}\mathcal{D}[b_2]\rho=(\gamma_{2,\text{int}}/2)(2b_2\rho b_2^\dagger
-\rho b_2^\dagger b_2-b_2^\dagger b_2\rho)$
to the master 
equation~\eqref{master_equation}. An imperfect JQF provides a 
significant reduction of the Purcell decay, even if it does not make it 
vanish completely. The internal decay of the qubit is modeled by adding 
$\gamma_\text{q,int}\mathcal{D}[b_1]\rho$ to the master equation, and 
Fig.~\ref{fig_decay_imperfections}(c) shows the influence of the different 
$T_{1,\text{q,int}}=1/\gamma_{\text{q,int}}$. Even for 
$T_{1,\text{q,int}}=50\text{ }\mu\text{s}$~\cite{burnett_npjqi19}, the 
difference in the error probability $1-F$ is significant, becoming more than an order of 
magnitude for
$T_{1,\text{q,int}}=500\text{ }\mu\text{s}$~\cite{place_ncomms21,wang_npjqi22}. 
These differences could be larger if the Purcell limited $T_1$ time, 
$1/\kappa_\text{Purcell}\approx 33\text{ }\mu\text{s}$, were lower, e.g., by choosing a 
smaller detuning $\omega_\text{r}-\omega_{\text{t},1}$, which may decrease the 
gate time.

In App.~\ref{App:dde_decay}, we check the
approximation~\eqref{free_evolution_appr} used in the derivation of 
the master equation~\eqref{master_equation} numerically. Without this
approximation,
delay differential equations are obtained, and they can be
solved for the single-excitation subspace. The differences between the two 
models can only be seen with significant zoom factors, with the $1-F$ curves 
deviating on the order of $10^{-6}$ or less.

\section{Measurement}\label{sec_measurement}

The dispersive shifts of the resonator frequency depend on the state of the 
transmon~\cite{gambetta_notes13}, and this is the standard physical mechanism for 
the qubit measurement in the superconducting quantum 
processors~\cite{arute_nature19,jurcevic_qst21,gong_science21,zhu_sb22,zhang_sciadv22,
jeffrey_prl14}. The desirable parameter regime is where the internal losses 
are negligible on the time scales of the duration of the measurement. Here, we only consider zero internal losses in the model, but these 
losses could be added as in the previous section. Since 
the considered setup [Fig.~\ref{fig_filters}(d)] is in the reflection geometry 
and under the assumption of zero internal losses, all of the incident
radiation gets reflected due to the energy conservation. Hence, no 
information can be gained from the amplitude of the reflected field, and only 
its phase carries the information about the qubit.
The experimentally accessible $I$~and~$Q$ values could be obtained by the
phase-preserving amplification~\cite{clerk_rmp10} and mixing with a local 
carrier on an $IQ$-mixer~\cite{arute_nature19}. 
Then the $I$~and~$Q$ values are proportional to the sine and cosine of the 
reflected phase~\cite{naghiloo_2019}. With the fast analog to digital converters~\cite{stefanazzi_rsi22,tholen_rsi22}, 
an $IQ$-mixer may not be needed, and then the relationship between the reflected 
phase and the final processed values could in principle be arbitrary.

\begin{figure}[t]
\begin{center}
\includegraphics{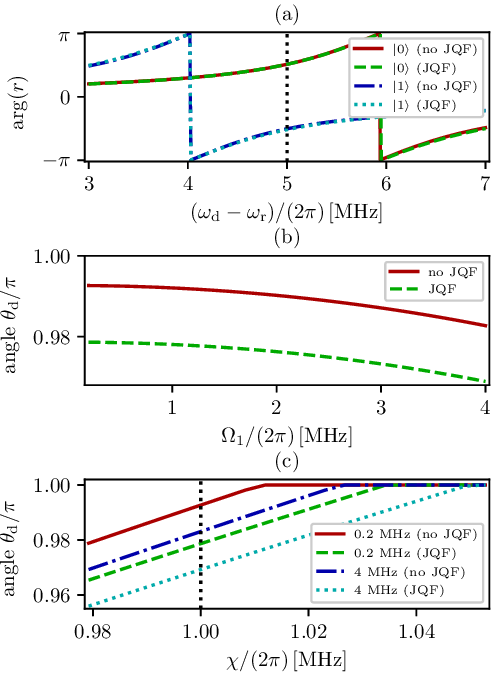}
\end{center}
\caption{(a) The phase of the reflection 
coefficient~\eqref{reflection_coefficient} as a function of the probe 
frequency~$\omega_\text{d}$ for the Rabi frequency 
$\Omega_1/(2\pi)=4\text{ MHz}$, corresponding to $-125$~dBm [cf. 
Eq.~\eqref{Omega_m_def}] and $16$~photons on average 
inside an empty resonator that is probed on resonance 
($\omega_\text{d}=\omega_\text{r}$) without a JQF in front. In this and 
the other subfigures, the qubit is 
initialized either in the state~$|0\rangle$ or the state~$|1\rangle$ and evolved for a 
fixed time $t_\text{f}=20/\kappa\approx 1.6\text{ }\mu\text{s}$. (b) The 
angle~\eqref{cos_theta_d_def} between the states $|0\rangle$ and $|1\rangle$ 
as a function of the Rabi frequency $\Omega_1$ at the probing frequency 
$(\omega_\text{d}-\omega_\text{r})/(2\pi)=5\text{ MHz}$ (vertical dotted 
black line in (a)). (c) The 
angle~\eqref{cos_theta_d_def} 
as a function of the dispersive shift $\chi$ (related to the coupling 
$g_\text{r}$ by Eq.~\eqref{chi_definition}) for a
probing frequency $\omega_\text{d}$ optimized numerically~\cite{powell_2009,nlopt} to give the maximum 
angle. The legend indicates the values of $\Omega_1/(2\pi)$. The 
vertical dotted line indicates $\chi/(2\pi)=1\text{ MHz}$ ($g_\text{r}/(2\pi)=109.544\text{ MHz}$) used 
in (a) and (b), and also all the other figures of this article. This value satisfies the condition $\chi=\kappa/2$.
\label{fig_reflection}}
\end{figure}

Since the JQF is far detuned from the probe (around $2$~GHz in our assumed parameters), it is weakly excited even for moderate powers of the 
probing field. E.g., the maximum JQF population 
$\langle b_2^\dagger b_2\rangle$ for the parameters of 
Fig.~\ref{fig_reflection} is around 
$3\times 10^{-4}$. Therefore, the noise contribution is assumed 
to be negligible, and instead of the more advanced theoretical descriptions of the 
measurement mechanism that also include the noise contributions~\cite{gambetta_pra07,wiseman_milburn_2009}, we 
take the simple 
approach of only considering the expectation values of the complex reflection 
coefficient $r$. The optimal situation is when the reflection coefficients, 
interpreted as 2D vectors with the components $\text{Re}[r]$ and 
$\text{Im}[r]$, point in the opposite directions for
the computational basis states $|0\rangle$~and~$|1\rangle$~\cite{sank_phd_thesis}.
In the dispersive approximation, this gives the condition $\chi=\kappa/2$, 
so that the dispersive shift $\chi$ is large enough compared to the resonator 
linewidth $\kappa$ to obtain the maximum angle of $\pi$ between the reflection 
coefficients.

In App.~\ref{App:reflection_coefficient}, we derive the expression for the 
reflection coefficient
\begin{gather}\label{reflection_coefficient}
\begin{aligned}
&r=1
-i\sum_{m=1}^2\sum_{j,j'}
\frac{\omega_{m,j'j}}{\sqrt{\omega_1\omega_m}}\frac{\sqrt{\Gamma_1\Gamma_m}}{\Omega_1}
C_{m,jj'}\\
&\times\tr_\text{s}[\sigma_{m,jj'}\tilde{\rho}_\text{s}] \cos(k_{\omega_\text{d}}x_m)
\end{aligned}
\end{gather}
at the position $x_2^+=x_2+\epsilon$ for $\epsilon\rightarrow 0$ from 
above, i.e., just to the right of the last subsystem attached to the 
transmission line (the JQF). The expression is written in terms of the reference Rabi 
frequency $\Omega_1$, and the overall propagation phase 
$e^{2ik_{\omega_\text{d}}x_2^+}$ has been removed. The master 
equation~\eqref{master_equation} is solved to evaluate the expectation values
$\tr_\text{s}[\sigma_{m,jj'}\tilde{\rho}_\text{s}]$, and the reflection
coefficient as a function of time is calculated.
The true steady-state reflection coefficient for both of the two initial qubit 
states, $|0\rangle$ and $|1\rangle$, is the same since the state~$|1\rangle$ 
eventually decays into the state~$|0\rangle$. Therefore, we evolve the master 
equation for a finite time, which is long enough for the transients to 
disappear, but short enough that the state~$|1\rangle$ does not 
decay significantly. We choose the evolution time 
$t_\text{f}=20/\kappa\approx 1.6\text{ }\mu\text{s}$ in 
Fig.~\ref{fig_reflection}.

In Fig.~\ref{fig_reflection}(a), the arguments of the complex reflection 
coefficients $\arg(r)$ are shown as a function of probe frequency. For the 
probe frequency in the middle, 
$(\omega_\text{d}-\omega_\text{r})/(2\pi)=5\text{ MHz}$, close to the 
desired angle of $\pi$ between the states $|0\rangle$ and $|1\rangle$ is 
obtained. To quantify how close the angle is to $\pi$, we calculate the 
smallest angle~$\theta_\text{d}$ between the two complex reflection 
coefficients via the dot product,
\begin{gather}\label{cos_theta_d_def}
\cos\theta_\text{d} = \frac{\text{Re}[r_{|0\rangle}]\text{Re}[r_{|1\rangle}]
+\text{Im}[r_{|0\rangle}]\text{Im}[r_{|1\rangle}]
}{|r_{|0\rangle}||r_{|1\rangle}|},
\end{gather}
where $r_{|0\rangle}$ and $r_{|1\rangle}$ are the reflection coefficients for 
the initial states $|0\rangle$ and $|1\rangle$, respectively. In 
Fig.~\ref{fig_reflection}(b), $\theta_\text{d}$ is shown as a function of the 
probe Rabi frequency $\Omega_1$.

Due to the transmon and resonator subsystem becoming more nonlinear for larger probe 
Rabi frequencies, the angle $\theta_\text{d}$ decreases. Addition of the JQF also 
decreases $\theta_\text{d}$, although by a small fixed amount, about $1.5\%$.
Despite choosing the parameters such that $\chi=\kappa/2$, the 
angle $\theta_\text{d}$ does not reach $\pi$ even for a weak probe and without a JQF. 
This could be caused by the fact that the Schrieffer-Wolff transformation used 
to obtain the expression~\eqref{chi_definition} for $\chi$ is a perturbative method and 
hence inexact. Other reasons for the discrepancy could be that our model does 
not make the 
dispersive approximation, under which the condition $\chi=\kappa/2$ is derived, 
and because there is some uncertainty with the heuristic procedure of choosing 
the finite evolution time $t_\text{f}$. We have verified that 
increasing $\chi$ (by increasing the coupling $g_\text{r}$) slightly is 
sufficient to reach the maximum angle of $\pi$, as shown in 
Fig.~\ref{fig_reflection}(c). In this subfigure, the angle $\theta_\text{d}$ 
is plotted as a function of the coupling $\chi$ for an optimal $\omega_\text{d}$ 
(obtained by the numerical optimization~\cite{powell_2009,nlopt}). Once $\chi$ is large enough to 
obtain $\theta_\text{d}=\pi$ for a chosen probe power, the measurement is not 
expected to improve, but a larger coupling might still be useful for 
decreasing the gate time.

The above results suggest that the JQFs have a very similar behavior to the 
unsaturable band-rejection Purcell filters~\cite{reed_apl10}, in that each 
Purcell filter acts as a far off-resonant scatterer during the measurement, 
adding a small phase shift to the reflected field. Hence, if JQFs are combined 
with the frequency multiplexing, we expect the angle decrease between the 
computational basis states for each qubit to be small, as long as the number of the 
qubits in each multiplexed group is not much bigger than currently used 
(around $6$~\cite{arute_nature19}). Other considerations for the frequency 
multiplexing that are not included in our model, such as the performance of 
the quantum limited amplifiers~\cite{mutus_apl14}, are expected to play a 
much bigger role than the presence of the JQFs.

\section{Control}\label{sec_control}

To verify the controllability of the system despite the complications arising from 
coupling to the qubit through the JQF and the resonator, we show that the Pauli $\sigma_x$ 
gate can be implemented with 
high fidelity. For a two-level atom with a directly attached control line, 
this can be accomplished with a simple rectangular pulse. The setup where the 
two-level atom is replaced with a transmon and JQF added in the control 
line [Fig.\ref{fig_filters}(c)] requires pulses that are more carefully 
chosen~\cite{kono_ncomms20,masuda_njp21}. For the setup that we are 
considering here [Fig.\ref{fig_filters}(d)], we use both the relatively simple 
pulses similar to Refs.~\cite{kono_ncomms20,masuda_njp21,motzoi_prl09} that do 
not require extensive calibration, and the more general Fourier series pulses 
inspired by Refs.~\cite{doria_prl11,motzoi_pra11} that achieve a
larger gate fidelity.

We maximize the average gate fidelity. For a
qubit, it is sufficient to average over the initial states at 6 cardinal 
points of the Bloch sphere, i.e., the eigenstates of the 3 Pauli 
matrices~\cite{bowdrey_pla02}. More efficiently, the averaging could be done 
by propagating the Pauli matrices themselves with the master 
equation, even though these matrices are not valid states~\cite{bowdrey_pla02}.
The system under consideration has more than two levels, but the states before 
and after a gate are mostly restricted 
to the qubit subspace. 
The leakage outside of the qubit subspace is accounted for by adding 
the states of the coupled transmon and resonator subsystem that also include the 
second excited state of the transmon and 4 excitations of the combined system in 
total. There are $12$ such states. Together with $11$ states of the JQF, this 
sets the total Hilbert space basis size of $132$. In principle, more general 
expressions for the average fidelity need to be used with a much larger set of 
operators being propagated by the master equation~\cite{nielsen_pla02}. To 
keep the simulation run time manageable, we use the expressions in 
Ref.~\cite{bowdrey_pla02} instead.

The expression for the average fidelity for the ideal operator $U=\sigma_x$ 
and the real superoperator $\mathcal{M}$ that we use is thus 
\begin{gather}\label{F_tilde_average_def}
\tilde{F}_\text{average}=\frac{1}{4}\tilde{F}(I)+\frac{1}{12}\sum_{j=x,y,z}\tilde{F}(\sigma_j),
\end{gather}
where
\begin{gather}\label{F_tilde_def}
\tilde{F}(A)=\tr_\text{s}[UAU^\dagger\mathcal{M}(A)].
\end{gather}
The Pauli matrices $\sigma_j$ and the identity operator $I$ in 
Eq.~\eqref{F_tilde_average_def} are interpreted as operators 
on the entire ($132$-dimensional) Hilbert space but only have non-zero matrix 
elements for the qubit subspace. In Eq.~\eqref{F_tilde_def}, $\mathcal{M}(A)$ is calculated by initializing the master equation with 
the operator $A$ instead of the initial density matrix and propagating until 
the final time $t_\text{f}$. It is not possible to simplify  
$\tilde{F}(I)$ to a constant like in Ref.~\cite{bowdrey_pla02}, 
since $I$ is not an identity operator on the entire Hilbert space.
Compared to the fidelity $F$ given by Eq.~\eqref{F_def}, the trace over the JQF 
is not performed in Eq.~\eqref{F_tilde_def}, requiring the JQF to be in the ground state~$|0_2\rangle$ at the 
end of the gate. This ensures that the JQF does not disturb the qubit by 
emitting a photon after the gate is performed.

\begin{figure}[t]
\begin{center}
\includegraphics{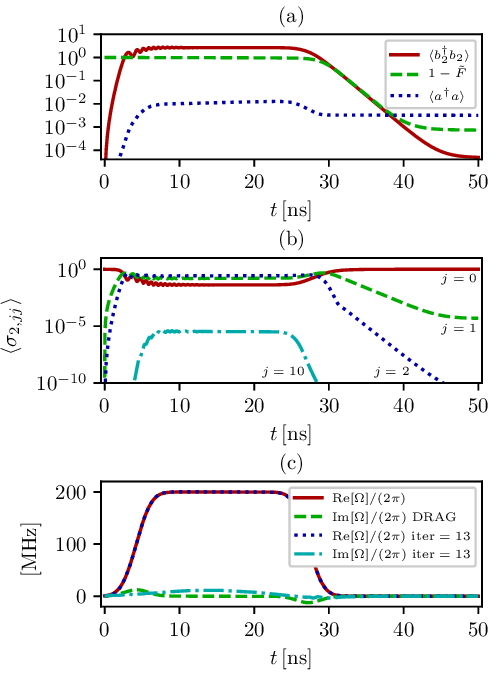}
\end{center}
\caption{The result of the simpler pulse shape [cf. Eq.~\eqref{f_simple_def}]
optimization with a DRAG~\cite{motzoi_prl09} correction. (a) The error probability $1-\tilde{F}(|0\rangle\langle 0|)$ where 
$\tilde{F}(A)$ 
is given by Eq.~\eqref{F_tilde_def}, the resonator
population 
$\langle a^\dagger a\rangle$, and the JQF population $\langle b_2^\dagger b_2\rangle$
as functions of time for the setup initialized in the state~$|0\rangle$ and 
driven to the state $\sigma_x|0\rangle=|1\rangle$ by the time-dependent Rabi 
frequency shown in (c) by the solid red ($\text{Re}[\Omega](t)$) and dashed 
green ($\text{Im}[\Omega](t)$) curves. In Eq.~\eqref{f_simple_def}, we set $\Omega_\text{max}/(2\pi)=200\text{ MHz}$, 
corresponding to $-91$ dBm [cf. Eq.~\eqref{Omega_m_def}], and 
$1/\sigma_\text{f}=\kappa/0.02=628\text{ MHz}$.
The DRAG corrections have the form: $\text{Im}[\Omega](t)=C_{\text{DRAG},\text{Im}}\od{}{t}\text{Re}[\Omega](t)$, and
$\Delta_\text{d}(t)=C_{\text{DRAG},\Delta}\text{Re}[\Omega]^2(t)$ (not shown) 
with $C_{\text{DRAG},\text{Im}}$ and $C_{\text{DRAG},\Delta}$ optimized numerically~\cite{powell_2009,nlopt}. (b) Populations of the individual JQF levels 
$\langle \sigma_{2,jj}\rangle$.
The achieved fidelity is $\tilde{F}(|0\rangle\langle 0|)=0.9993$ 
($\tilde{F}_\text{average}=0.9981$). The dotted blue ($\text{Re}[\Omega](t)$) 
and dash-dotted cyan ($\text{Im}[\Omega](t)$)
curves in (c) show the pulse shapes obtained after $13$ iterations of the optimal 
control algorithm described in the text, using the initial pulse shape with 
$\text{Re}[\Omega](t)$ given by the solid red curve and 
$\text{Im}[\Omega](t)=0$. The achieved fidelity after $13$ iterations is $\tilde{F}(|0\rangle\langle 0|)=0.9993$ ($\tilde{F}_\text{average}=0.9994$).
\label{fig_simple_gate}}
\end{figure}

For the simpler pulse shape, we choose a Gaussian-filtered rectangular pulse
\begin{gather}\label{f_simple_def}
\text{Re}[\Omega](t) = \frac{\Omega_\text{max}}{\sigma_\text{f}\sqrt{2\pi}}\int_{t_\text{start}}^{t_\text{end}}
\exp\left(-\frac{(t-t')^2}{2\sigma_\text{f}^2}\right)\dif t', 
\end{gather}
where $0\leq t_\text{start},t_\text{end}\leq t_\text{f}$. The above integral 
can be evaluated in terms of the error functions. The initial value 
$\text{Re}[\Omega](0)$ is never exactly zero, but $t_\text{start}$ can be 
chosen such that $\text{Re}[\Omega](0)$ is below a certain tolerance 
($\text{Re}[\Omega](0)/(2\pi) < 0.2\text{ MHz}$ in Fig.~\ref{fig_simple_gate}(c)). 
The carrier frequency of the drive is set equal to the qubit transition frequency,
i.e., $\omega_\text{d}=\omega_{1,10}$. Optionally, a 
correction is applied to the imaginary quadrature 
$\text{Im}[\Omega](t)=C_{\text{DRAG},\text{Im}}\od{}{t}\text{Re}[\Omega](t)$, 
and a power dependent frequency shift 
$\Delta_\text{d}(t)=C_{\text{DRAG},\Delta}\text{Re}[\Omega]^2(t)$ is used similar to the
Derivative Removal by Adiabatic Gate (DRAG)~\cite{motzoi_prl09}.

For a transmon without a JQF, the DRAG correction constants $C_{\text{DRAG},\text{Im}}$ 
and $C_{\text{DRAG},\Delta}$ have analytical expressions~\cite{motzoi_prl09}. For our setup with the 
JQF, we find that these constants need to be optimized numerically~\cite{powell_2009,nlopt} to 
yield any improvement for the fidelity. However, the improvement is so small as to be 
negligible. Without the DRAG correction ($C_{\text{DRAG},\text{Im}}=0$ 
and $C_{\text{DRAG},\Delta}=0$), we get $\tilde{F}_\text{average}=0.9980$. 
With the DRAG correction, $\tilde{F}_\text{average}=0.9981$. Some of the 
initial states achieve higher fidelities, as shown in 
Fig.~\ref{fig_simple_gate} with $\tilde{F}(|0\rangle\langle 0|)=0.9993$. Using 
the optimal control approach described below with only $13$ iterations, we see 
that a better correction $\text{Im}[\Omega](t)$ (the dash-dotted cyan curve in 
Fig.~\ref{fig_simple_gate}(c)) is not proportional to the time derivative of 
$\text{Re}[\Omega](t)$, contrary to DRAG. The optimal 
control with $13$ iterations achieves $\tilde{F}_\text{average}=0.9994$
($\tilde{F}(|0\rangle\langle 0|)=0.9993$) while keeping the pulse shapes simple.

To reach higher fidelities, we run the optimal control for more iterations.
We consider the Fourier series parametrization of the pulses with a
finite number of terms to 
limit the bandwidth. We can write this parametrization
\begin{subequations}
\begin{align}
&\text{Re}[\Omega](t)=\sqrt{\frac{2}{t_\text{f}}}\sum_{p=1}^{N_\text{c}} a_p \sin(\omega_p t),\displaybreak[0]\\
&\text{Im}[\Omega](t)=\sqrt{\frac{2}{t_\text{f}}}\sum_{p=1}^{N_\text{c}} b_p \sin(\omega_p t),
\end{align}
\label{Re_Im_Omega_parametrization}
\end{subequations}
where $N_\text{c}$ is the maximum number of the Fourier components, and
$\omega_p=p\pi/t_\text{f}$.
By construction, $\sin(0)=\sin(\omega_p t_\text{f})=0$.

\begin{figure}[t]
\begin{center}
\includegraphics{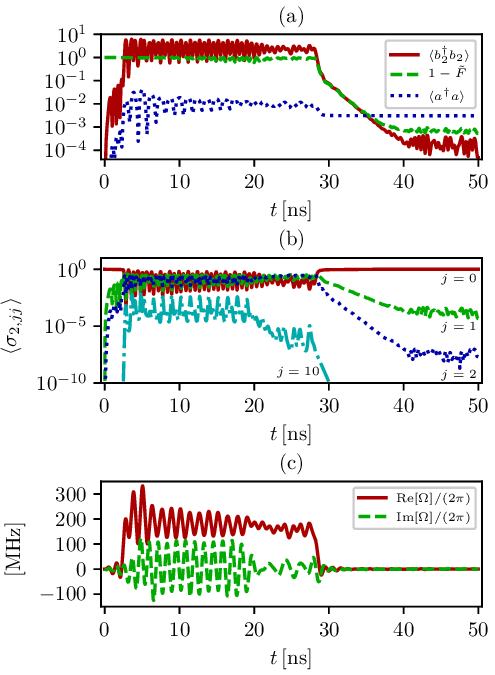}
\end{center}
\caption{The result of the optimal control. (a) The error probability $1-\tilde{F}(|0\rangle\langle 0|)$ where $\tilde{F}(A)$ 
is given by Eq.~\eqref{F_tilde_def}, the resonator
population 
$\langle a^\dagger a\rangle$, and the JQF population $\langle b_2^\dagger b_2\rangle$
as functions of time for the setup initialized in the state~$|0\rangle$ and 
driven to the state $\sigma_x|0\rangle=|1\rangle$ by the time-dependent Rabi 
frequency shown in (c). (b) Populations of the individual JQF levels 
$\langle \sigma_{2,jj}\rangle$.
The achieved fidelity is $\tilde{F}(|0\rangle\langle 0|)=0.9995$ ($\tilde{F}_\text{average}=0.9996$).
The number of the Fourier components in the 
parametrization~\eqref{Re_Im_Omega_parametrization} is $N_\text{c}=100$ (for 
each of $\text{Re}[\Omega]$ and $\text{Im}[\Omega]$). This corresponds to the 
highest Fourier frequency $\omega_{N_\text{c}}/(2\pi)=1\text{ GHz}$.
\label{fig_gate}}
\end{figure}

Optimizing a function with many variables (we choose $2N_\text{c}=200$) is faster if a gradient-based 
algorithm is used. We use the reverse mode automatic differentiation, as 
explained in App.~\ref{App:optimal_control}, to 
calculate the gradient, which is used in the LBFGS algorithm~\cite{nocedal_moc80,nlopt} to find 
the maximum of $\tilde{F}_\text{average}$.
For the 
reverse mode automatic differentiation, the cost of calculating the gradient is 
independent of the number of the variables $2N_\text{c}$. Using the 4$^\text{th}$ 
order Runge-Kutta method for the propagation of the master equation, the time 
to calculate the 
fidelity and the gradient is around $8$ times larger than the 
time to calculate the fidelity without the gradient.

For the initial qubit state~$|0\rangle$, the resulting evolutions of the error probability $1-\tilde{F}(|0\rangle\langle 0|)$ and the populations of the resonator and 
the JQF are shown in Fig.~\ref{fig_gate}(a). The corresponding
time-dependent Rabi frequency is shown in Fig.~\ref{fig_gate}(c). The 
optimization has not finished after $2000$ iterations 
taking around $50$ days with the convergence rate becoming extremely small. 
It was run on a machine with a
Ryzen~3700X CPU and a Radeon~VII GPU, with the most computationally expensive 
part, the sparse matrix-vector multiplications, being performed on the GPU. A pure CPU calculation 
is about $4.5$ times slower. Both implementations could be 
optimized further to speed up the calculations, and there may be a more 
efficient parametrization the the pulse shapes. Contrary to the setups without 
the JQF, we cannot perform the 
evolution in a closed system (with a Schr{\"o}dinger equation) while 
optimizing the control pulse shapes~\cite{motzoi_prl09}, with the master 
equation only being used to evaluate the final pulses. The JQF has a short 
life time of $1/\gamma_2=1.6\text{ ns}$ by design, which is much smaller than the gate time. Hence, a 
slower master equation evolution is needed to accurately simulate the dynamics 
also during the optimization. 

The populations of the individual levels of the JQF are shown in 
Figs.~\ref{fig_simple_gate}(b)~and~\ref{fig_gate}(b). In both cases, the JQF is driven to the higher excitation 
levels by the strong control fields, effectively decoupling it from 
the transmission line and permitting control of the qubit. While the simpler 
pulses just reach some steady state level of the JQF population, the pulses 
found by optimal control induce fast oscillations, which result in 
a higher gate fidelity $\tilde{F}_\text{average}=0.9996$ 
($\tilde{F}(|0\rangle\langle 0|)=0.9995$). For the optimal control 
parametrization~\eqref{Re_Im_Omega_parametrization}, we use 
$N_\text{c}=100$, giving the highest Fourier frequency 
$\omega_{N_\text{c}}/(2\pi)=1\text{ GHz}$.
Since it is 
possible to synthesize microwave pulses with bandwidths of several 
GHz~\cite{kalfus_ieeetqe20,stefanazzi_rsi22,tholen_rsi22}, the pulse shapes in Fig.~\ref{fig_gate}(c) are feasible. 

Combination of the JQFs with the frequency multiplexing where multiple qubits 
are controlled by the same transmission line will likely require tuning of the 
control pulse shapes to reach high gate fidelities, because several qubit transitions will be within the 
bandwidth of the pulses shown in 
Figs.~\ref{fig_simple_gate}(c)~and~\ref{fig_gate}(c). Ideally, this would be 
accomplished by repeating the pulse shape optimizations using a Hilbert 
space that includes the entire system of several transmons and resonators. If 
the Hilbert space dimension $132$ is used for each set of two transmons and a 
resonator that stores an filters one qubit
(like we do for the simulations in this section), then the 
total Hilbert space dimension is $132^{N_\text{q}}$
with, e.g., ${N_\text{q}}=6$~\cite{arute_nature19}. This makes the storage 
requirements for the density matrix prohibitive. It may be possible to side 
step this problem by using optimal control with the stochastic wave 
functions~\cite{abdelhafez_pra19} or tensor networks~\cite{doria_prl11}. Using 
the experimental setup directly is also an option~\cite{werninghaus_npjqi21}.

\section{Conclusion}\label{sec_conclusion}

We have shown theoretically that it is possible to construct a saturable 
Purcell filter using an artificial atom directly attached to the transmission 
line. This filter suppresses the Purcell decay when the control fields are 
absent and can be effectively switched off by saturation when the control 
fields are present. This allows both the control and measurement of the qubit
to be performed using a single transmission line while 
maintaining long coherence time. Our results can be used to decrease the 
number of the needed transmission lines in the superconducting quantum 
processors and other setups involving superconducting artificial atoms. Further 
reductions in the number of the transmission lines could be achieved by 
combining saturable Purcell filters with the frequency multiplexing.

\begin{acknowledgments}
The authors acknowledge T. Shitara, S. Goto, and Y. Sunada for fruitful discussions. This 
work was supported by JST ERATO (Grant No. JPMJER1601), JST Moonshot R\&D 
(Grant Nos. JPMJMS2067-3 and JPMJMS2061-2-1-2), and JSPS KAKENHI (Grant No. 
22K03494).
\end{acknowledgments}

\appendix

\section{Derivation of the master equation}\label{App:master_equation}
In this appendix, we derive the master equation~\eqref{master_equation}, 
following 
Refs.~\cite{koshino_prapplied20,kono_ncomms20,ott_pra13,lehmberg_pra70}. The 
derivation is for $N$ attached subsystems, i.e., not limited to $N=2$ as in 
the main text. In this case, the summation in Hamiltonian given by 
Eq.~\eqref{H_i} and the following ones is to $N$ instead of $2$; and 
Eqs.~\eqref{Gamma_m_def}, \eqref{omega_m_def}, and \eqref{O_m_def} need to be
redefined depending on the attached subsystems. Then the Heisenberg equations 
of motion for the field operators are
\begin{align}\label{c_Heisenberg_equation}
&\dot{c}_\omega = -i\omega c_\omega 
- i\sum_{m=1}^{N} g_m(\omega)(\mathcal{O}_m-\mathcal{O}_m^\dagger)
\end{align}
with the solutions
\begin{gather}\label{c_Heisenberg_equation_solution}
\begin{aligned}
&c_\omega 
=c_\omega(0)e^{-i\omega t}
- i\sum_{m=1}^{N} g_m(\omega)\\
&\times\int_0^t  \Big(\mathcal{O}_m(t-t')-\mathcal{O}_m^\dagger(t-t')\Big)
e^{-i\omega t'}\dif t'.
\end{aligned}
\end{gather}
Here and below, the indication of the time-dependence of the operators is omitted 
for brevity as long as it is of the simple form: $c_\omega$ means $c_\omega(t)$ in the above expressions.
Note that the term involving $\mathcal{O}_m^\dagger$ is present, 
because the rotating wave approximation is \emph{not} 
performed~\cite{ott_pra13} in the Hamiltonian~\eqref{H_i}.
Once we calculate
\begin{gather}\label{c_plus_c_dagger}
\begin{aligned}
&c_\omega-c_\omega^\dagger 
= c_\omega(0)e^{-i\omega t}-c_\omega^\dagger(0)e^{i\omega t}
-i\sum_{m=1}^N g_m(\omega)\\
&\times\Bigg(\int_0^t\mathcal{O}_m(t-t')
\left(e^{-i\omega t'}-e^{i\omega t'}\right)\dif t'\\
&-\int_0^t\mathcal{O}_m^\dagger(t-t')
\left(e^{-i\omega t'}-e^{i\omega t'}\right)\dif t'\Bigg),
\end{aligned}
\end{gather}
we see that terms arising from not performing the rotating wave 
approximation appear as additional $e^{\pm i\omega t'}$ in the inner parentheses. 
These will allow us to extend the integrations over $\omega$ to the 
entire real line.

The expression for $c_\omega-c_\omega^\dagger$ is needed in the Heisenberg equation of motion for an arbitrary system operator~$Q$,
\begin{align}\label{Q_equation}
\begin{aligned}
&\dot{Q}=\frac{i}{\hbar}[H_\text{s},Q]+i\sum_{m=1}^N\int_0^\infty  g_m(\omega)\\
&\times\del{[\mathcal{O}_m^\dagger,Q] (c_\omega-c_\omega^\dagger) - (c_\omega-c_\omega^\dagger) [\mathcal{O}_m,Q]}\dif \omega,
\end{aligned}
\end{align}
written in the normal ordered form.
When we insert Eq.~\eqref{c_plus_c_dagger}, the 
normal ordering becomes important. The 
integrations over $\omega$ are carried out first. Using the expression for the 
coupling $g_m(\omega)=G_m\sqrt{\omega}\cos(k_\omega x_m)$, we have
\begin{gather}\label{delta_prime_integral}
\begin{aligned}
&\int_0^\infty g_m(\omega)g_n(\omega)\left(e^{-i\omega t'}-e^{i\omega 
t'}\right)\dif \omega\\
&=\frac{i\pi G_m G_n}{2}\Bigg(
\dot{\delta}\bigg(t'-\frac{x_m-x_n}{v_\text{g}}\bigg)
+\dot{\delta}\bigg(t'+\frac{x_m-x_n}{v_\text{g}}\bigg)\\
&+\dot{\delta}\bigg(t'-\frac{x_m+x_n}{v_\text{g}}\bigg)
+\dot{\delta}\bigg(t'+\frac{x_m+x_n}{v_\text{g}}\bigg)\Bigg),
\end{aligned}
\end{gather}
where $\dot{\delta}$ is the derivative of the Dirac delta function that has the 
property
\begin{gather}\label{delta_dot_property}
\int_{0}^{t} \dot{\delta}(t'-t_x) f(t-t') \dif t' = \dot{f}(t-t_x),
\end{gather}
as long as $0<t_x<t$, $f$ is an arbitrary (operator-valued) function, and the 
integration limits are chosen as to be relevant to the present derivation. The 
case with $t_x=0$ is defined with $\dot{f}(t)/2$ on the right hand side of 
Eq.~\eqref{delta_dot_property}. Since $t_x=0$ is equivalent to 
$x_m, x_n=0$ in Eq.~\eqref{delta_prime_integral}, this case could also be addressed by
setting $x_m, x_n=0$ in the integral on the left hand side of
Eq.~\eqref{delta_prime_integral}. Thus,
\begin{gather}\label{delta_dot_property_specific}
\begin{aligned}
&\int_{0}^{t} \Bigg(\dot{\delta}\bigg(t'-\frac{x_m\pm x_n}{v_\text{g}}\bigg)
+\dot{\delta}\bigg(t'+\frac{x_m\pm x_n}{v_\text{g}}\bigg)\Bigg)\\
&\times \mathcal{O}_n(t-t')\dif t'\\
& = \dot{\mathcal{O}}_n\bigg(t-\frac{|x_m\pm x_n|}{v_\text{g}}\bigg)
\theta_\text{H}\bigg(t-\frac{|x_m\pm x_n|}{v_\text{g}}\bigg),
\end{aligned}
\end{gather}
where $\theta_\text{H}$ is the Heaviside theta function.

The integral in Eq.~\eqref{delta_prime_integral} can also be evaluated with 
$g_m(\omega)=G_m(\sqrt{\omega}/\sqrt{1+\mathcal{A}\omega^2})\cos(k_\omega x_m)$ 
for $\mathcal{A}>0$, resulting in
\begin{gather}\label{delta_prime_integral_cutoff}
\begin{aligned}
&G_mG_n\int_0^\infty \frac{\omega\cos(k_\omega x_m)\cos(k_\omega x_n)}{1+\mathcal{A}\omega^2}\left(e^{-i\omega t'}-e^{i\omega 
t'}\right)\dif \omega\\
&=\Bigg(
\mathcal{K}\bigg(t'-\frac{x_m-x_n}{v_\text{g}}\bigg)
+\mathcal{K}\bigg(t'+\frac{x_m-x_n}{v_\text{g}}\bigg)\\
&+\mathcal{K}\bigg(t'-\frac{x_m+x_n}{v_\text{g}}\bigg)
+\mathcal{K}\bigg(t'+\frac{x_m+x_n}{v_\text{g}}\bigg)\Bigg),
\end{aligned}
\end{gather}
with
\begin{gather}\label{delta_prime_integral_cutoff_kernel}
\mathcal{K}(t)=-\frac{i\pi G_m G_n}{4\mathcal{A}}\text{sgn}(t)e^{-|t|/\sqrt{\mathcal{A}}},
\end{gather}
and $\text{sgn}$ being the sign function, $\text{sgn}(t)=2\theta_\text{H}(t)-1$.
In this case, the time integral corresponding to 
Eq.~\eqref{delta_dot_property_specific} cannot be evaluated explicitly. Since 
it is in a form of a convolution, the Laplace transform could be 
used~\cite{wodkiewicz_ap76}, but there is no algorithm for the numerical inverse 
Laplace transform that can work in all cases, necessitating selection 
among the different available 
algorithms~\cite{davies_jcp79,duffy_tms93,kuhlman_na13}. For simplicity, we 
find the Markovian master equation using a different integration order (first 
$t$ then $\omega$ instead of first $\omega$ then $t$) and using the 
approximation~\eqref{free_evolution_appr}~or~\eqref{driven_evolution_appr} 
from the beginning, as detailed in App.~\ref{App:cutoff_in_coupling}.

For the rest of this appendix, we continue with the coupling
$g_m(\omega)=G_m\sqrt{\omega}\cos(k_\omega x_m)$. Defining the noise operator
\begin{gather}\label{N_operator_def}
\mathcal{N}_m=\int_0^\infty  g_m(\omega)c_\omega(0) e^{-i\omega t}\dif \omega,
\end{gather}
and applying the rotating wave approximation, we can write
\begin{gather}\label{Q_equation2}
\begin{aligned}
&\dot{Q}=\frac{i}{\hbar}[H_\text{s},Q]
+i\sum_{m=1}^N\del{[\mathcal{O}_m^\dagger,Q]\mathcal{N}_m 
+ \mathcal{N}_m^\dagger [\mathcal{O}_m,Q]}\\
&+\sum_{m,n=1}^N \frac{i\pi G_m G_n}{2}[\mathcal{O}_m^\dagger,Q] 
\Bigg(\dot{\mathcal{O}}_n\bigg(t-\frac{|x_m - x_n|}{v_\text{g}}\bigg)\\
&+\dot{\mathcal{O}}_n\bigg(t-\frac{|x_m + x_n|}{v_\text{g}}\bigg)\Bigg)\\
&+\sum_{m,n=1}^N\frac{i\pi G_m G_n}{2}\Bigg(
\dot{\mathcal{O}}_n^\dagger\bigg(t-\frac{|x_m - x_n|}{v_\text{g}}\bigg)\\
&+\dot{\mathcal{O}}_n^\dagger\bigg(t-\frac{|x_m + x_n|}{v_\text{g}}\bigg)\Bigg)
[\mathcal{O}_m,Q],
\end{aligned}
\end{gather}
where the Heaviside theta function factors resulting from 
Eq.~\eqref{delta_dot_property_specific} are implicit. We make one more 
approximation by setting 
\begin{gather}\label{dot_O_approx}
\dot{\mathcal{O}}_n(t-t_x)
\approx-i\sum_{j,j'}\omega_{n,j'j}C_{n,jj'}\sigma_{n,jj'}(t-t_x),
\end{gather}
where $\omega_{n,j'j}=\omega_{n,j'}-\omega_{n,j}$.
This approximation can be viewed as applying Eq.~\eqref{Q_equation2} and ignoring all the 
terms besides $\frac{i}{\hbar}[H_\text{s},\mathcal{O}_n]$ due to the fact that 
the absolute frequencies $\omega_{n,j'j}$ are large compared to the couplings 
$G_n$. Thus, this is also a form of a rotating wave approximation.

In App.~\ref{App:dde_decay}, the equations of 
motion in the single-excitation subspace are derived from 
Eq.~\eqref{Q_equation2} without any further approximations besides Eq.~\eqref{dot_O_approx}. For the master equation~\eqref{master_equation}, Eq.~\eqref{Q_equation2} needs 
to be approximated such that it becomes local in time, i.e., does not contain
operators at the previous times $t-|x_m\pm x_n|/v_\text{g}$. We use either
approximation~\eqref{free_evolution_appr}~or~\eqref{driven_evolution_appr}
together with the approximation~\eqref{dot_O_approx}.
Additionally, we assume that the size of the ensemble is small, i.e.,
$|x_m\pm x_n|/v_\text{g}$ is short compared to the time scales of interest, 
and hence we set $\theta_\text{H}(t-|x_m\pm x_n|/v_\text{g})=1$ for all~$t$. 

Using the approximation~\eqref{free_evolution_appr}, inserting $G_m=\sqrt{\Gamma_m/(2\pi\omega_m)}$, identifying 
$\xi_{mn,j'j}$ given by Eq.~\eqref{xi_def} and 
$\mathcal{O}_{mn}=\sum_{j,j'}\xi_{mn,j'j}C_{n,jj'}\sigma_{n,jj'}$, we get
\begin{gather}\label{Q_equation3}
\begin{aligned}
&\dot{Q}=\frac{i}{\hbar}[H_\text{s},Q]+i\sum_{m=1}^N\del{[\mathcal{O}_m^\dagger,Q]\mathcal{N}_m 
+ \mathcal{N}_m^\dagger [\mathcal{O}_m,Q]}\\
&+\frac{1}{2}\sum_{m,n=1}^N\bigg(
[\mathcal{O}_m^\dagger,Q] \mathcal{O}_{mn} - \mathcal{O}_{mn}^\dagger [\mathcal{O}_m,Q]\bigg).
\end{aligned}
\end{gather}
The above equation with the drive approximation~\eqref{driven_evolution_appr} is obtained by setting $k_{n,j'j}=k_{\omega_\text{d}}$ in 
Eq.~\eqref{xi_def}.

Since the expectation values are the same in the Heisenberg and 
Schr\"odinger pictures, we have
\begin{gather}
\langle Q \rangle = \tr_\text{s}\tr_\text{f}[Q\rho(0)]
= \tr_\text{s}[Q(0)\rho_\text{s}],
\end{gather}
where $\tr_\text{s}$ ($\tr_\text{f}$) is the trace over the system 
(field) degrees of freedom, and $\rho_\text{s}=\tr_\text{f}[\rho]$. Taking the time 
derivative, we get
\begin{gather}\label{schroedinger_heisenberg_relation}
\tr_\text{s}\tr_\text{f}[\dot{Q}\rho(0)]
= \tr_\text{s}[Q(0)\dot{\rho}_\text{s}].
\end{gather}
The procedure to obtain the master equation~\eqref{master_equation} starts with inserting 
Eq.~\eqref{Q_equation3} into the left hand side of 
Eq.~\eqref{schroedinger_heisenberg_relation}. The resulting expression can 
then be rewritten in the form $\tr_\text{s}[Q(0)B]$, where $B$ is some 
system operator expression. Using the 
right hand side of Eq.~\eqref{schroedinger_heisenberg_relation}, the master 
equation is obtained as $\dot{\rho}_\text{s}=B$.

For any system operator $A$,
\begin{gather}\label{Q_to_rho_identity}
\tr_\text{s}\tr_\text{f}[[A,Q]\rho(0)]
=-\tr_\text{s}\left[Q(0) [A(0), \rho_\text{s}]\right].
\end{gather}
In $\tr_\text{s}\tr_\text{f}[[\mathcal{O}_m^\dagger,Q] \mathcal{N}_m\rho(0)]$ and 
$\tr_\text{s}\tr_\text{f}[\mathcal{N}_m^\dagger[\mathcal{O}_m,Q]\rho(0)]$, 
we assume that $\rho(0)=\rho_\text{s}(0)\otimes \rho_\text{f}(0)$ with the 
state of the field $\rho_\text{f}(0)=|\{\alpha_\omega\}\rangle\langle \{\alpha_\omega\}|$
 being a multimode coherent state. This state could be written as the displaced vacuum 
state
$|\{\alpha_\omega\}\rangle=D(\{\alpha_\omega\})|\text{vac}\rangle$, where the displacement 
operator is
\begin{gather}
D(\{\alpha_\omega\})=\exp\left(\int_0^\infty (\alpha_\omega c_\omega^\dagger(0) - \alpha_\omega^* c_\omega(0))\dif \omega\right).
\end{gather}
We have
\begin{gather}\label{c_omega_action_on_alpha_omega}
c_\omega(0)|\{\alpha_\omega\}\rangle = \alpha_\omega|\{\alpha_\omega\}\rangle.
\end{gather}
To relate $\alpha_\omega$ to the photon flux $\dot{n}$, we define Fourier 
transformed operators~\cite{blow_pra90}
\begin{gather}\label{c_t_def}
c_t=\frac{1}{\sqrt{2\pi}}\int_{0}^\infty c_\omega(0)e^{-i\omega t}\dif \omega,
\end{gather}
and then the photon flux is $\dot{n}=\langle c_t^\dagger c_t\rangle=|\alpha_t|^2$, where
\begin{gather}\label{alpha_t_def}
\alpha_t=\frac{1}{\sqrt{2\pi}}\int_{0}^\infty \alpha_\omega e^{-i\omega t}\dif \omega.
\end{gather}

For the operators~\eqref{N_operator_def}, it holds that
\begin{gather}\label{N_m_action_on_alpha_omega}
\mathcal{N}_m|\{\alpha_\omega\}\rangle = \Omega_m e^{-i\omega_\text{d} t}|\{\alpha_\omega\}\rangle,
\end{gather}
where we have defined the Rabi frequency
\begin{gather}\label{Omega_m_def_general}
\Omega_m e^{-i\omega_\text{d} t}=\int_0^\infty g_m(\omega) \alpha_\omega 
e^{-i\omega t}\dif \omega.
\end{gather}
To go from Eq.~\eqref{Omega_m_def_general} to Eq.~\eqref{Omega_m_def}, a 
narrow-bandwidth approximation is made, 
$g_m(\omega)\approx g_m(\omega_\text{d})$, resulting in 
$\Omega_m e^{-i\omega_\text{d} t}=\sqrt{2\pi}g_m(\omega_\text{d})\alpha_t$. 
Since $\dot{n}=|\alpha_t|^2$, we set 
$\alpha_t=\sqrt{\dot{n}}e^{-i\omega_\text{d}t}e^{i\phi}$ for some phase $\phi$, 
and then Eq.~\eqref{Omega_m_def} is obtained. For the reflection coefficient 
calculations [Sec.~\ref{sec_measurement} and 
App.~\ref{App:reflection_coefficient}], we also need to consider the case of the 
infinitely narrow bandwidth continuous wave input, where
\begin{gather}\label{c_omega_action_on_ket_alpha}
c_{\omega}(0)|\{\alpha_\omega\}\rangle
=\sqrt{2\pi \dot{n}}e^{i\phi}\delta(\omega-\omega_\text{d})|\{\alpha_\omega\}\rangle.
\end{gather}

Using Eq.~\eqref{N_m_action_on_alpha_omega}, the terms $\tr_\text{s}\tr_\text{f}[[\mathcal{O}_m^\dagger,Q] \mathcal{N}_m \rho(0)]$ and 
$\tr_\text{s}\tr_\text{f}[\mathcal{N}_m^\dagger [\mathcal{O}_m,Q]\rho(0)]$ are simplified 
 into the form where Eq.~\eqref{Q_to_rho_identity} can be applied. These terms 
give rise to the drive 
Hamiltonian~\eqref{drive_Hamiltonian}.
Together with the other terms rewritten either using either 
Eq.~\eqref{Q_to_rho_identity} or in 
a similar way, we get the master equation
\begin{gather}\label{master_equation_original_frame}
\begin{aligned}
&\dot{\rho}_\text{s}
=-\frac{i}{\hbar}\left[H_\text{d}+H_\text{s},\rho_\text{s}\right]\\
&+\frac{1}{2}\sum_{m,n=1}^N\left(\mathcal{O}_{mn}\rho_\text{s}\mathcal{O}_m^\dagger-\mathcal{O}_m^\dagger\mathcal{O}_{mn}\rho_\text{s}\right)\\
&+\frac{1}{2}\sum_{m,n=1}^N\left(\mathcal{O}_n\rho_\text{s}\mathcal{O}_{nm}^\dagger-\rho_\text{s}\mathcal{O}_{nm}^\dagger\mathcal{O}_n\right).
\end{aligned}
\end{gather}
Transforming Eq.~\eqref{master_equation_original_frame} into the rotating frame with respect to the 
Hamiltonian~\eqref{H_0_def} by substituting $\rho_\text{s}=e^{-iH_0t/\hbar}\tilde{\rho}_\text{s}e^{iH_0t/\hbar}$ results in the master equation~\eqref{master_equation} with
\begin{gather}\label{Hamiltonian_rotating_frame_initial}
\tilde{H}=e^{iH_0 t/\hbar}\left(H_\text{d}+H_\text{s}\right)e^{-iH_0 t/\hbar}-H_0.
\end{gather}

Frequencies $\omega_{0,m,j}$ in 
Eq.~\eqref{H_0_def} are chosen such that
$\frac{i}{\hbar}[H_0,\mathcal{O}_m]=-i\omega_\text{d}\mathcal{O}_m$, and thus the factors 
$e^{\pm i\omega_\text{d} t}$ in Eq.~\eqref{drive_Hamiltonian}
are canceled. By 
inserting $\mathcal{O}_m=C_{m,jj'}\sigma_{m,jj'}$, the equivalent condition is
\begin{gather}\label{H_0_condition}
\omega_{0,m,j'}-\omega_{0,m,j}=\omega_\text{d}
\end{gather}
for every $C_{m,jj'}\neq 0$. It is possible to satisfy this condition, since 
$\mathcal{O}_m$ is an annihilation operator, and hence $C_{m,jj'}\neq 0$ only 
if $j$ and $j'$ correspond to the eigenstates with excitation numbers $N_{m,j}$ and 
$N_{m,j'}=N_{m,j}+1$, respectively. Thus, setting 
$\omega_{0,m,j}=N_{m,j}\omega_\text{d}$ satisfies Eq.~\eqref{H_0_condition}.  
With this choice, Eq.~\eqref{Hamiltonian_rotating_frame_initial} becomes 
Eq.~\eqref{Hamiltonian_rotating_frame}.

\section{Cutoff in the coupling}\label{App:cutoff_in_coupling}
In this appendix, we derive the master equation using the coupling between the 
subsystems and the transmission line
$g_m(\omega)=G_m(\sqrt{\omega}/\sqrt{1+\mathcal{A}\omega^2})\cos(k_\omega x_m)$ with 
$\mathcal{A}>0$~\cite{bamba_pra14,malekakhlagh_pra16,malekakhlagh_prl17,gely_pra17} instead of $g_m(\omega)=G_m\sqrt{\omega}\cos(k_\omega x_m)$ 
that was used in all of the main text and the other appendices. Similar to 
App.~\ref{App:master_equation}, $N$ attached subsystems are considered, 
instead of setting $N=2$ as in the main text. The discussion of 
the challenges associated with not using any approximations can be found below 
Eq.~\eqref{delta_prime_integral_cutoff} in App.~\ref{App:master_equation}. For 
simplicity, we use the approximation~\eqref{free_evolution_appr} from the 
beginning (using the approximation \eqref{driven_evolution_appr} is 
accomplished by replacing the frequencies). Inserting
$\mathcal{O}_{m}=\sum_{j,j'}C_{m,jj'}\sigma_{m,jj'}$ into 
Eq.~\eqref{c_Heisenberg_equation_solution}, applying the 
approximation~\eqref{free_evolution_appr}, and setting $\omega_{m,j'j}=\omega_{m,j'}-\omega_{m,j}$, gives
\begin{gather}
\begin{aligned}
&c_\omega 
=c_\omega(0)e^{-i\omega t}\\
&- i\sum_{m=1}^{N} \sum_{j,j'}g_m(\omega)
\int_0^t \Big(C_{m,jj'}\sigma_{m,jj'}e^{-i(\omega-\omega_{m,j'j})t'}\\
&-C_{m,jj'}^*\sigma_{m,j'j}e^{-i(\omega+\omega_{m,j'j})t'}\Big)\dif t'.
\end{aligned}
\end{gather}
Following Ref.~\cite{lehmberg_pra70}, the time integral could be approximated 
by extending the upper limit to infinity and using the identity (related to 
the Sokhotski-Plemelj theorem)
\begin{align}
\int_0^\infty e^{\pm i\epsilon s} \dif s = \pi\delta(\epsilon) \pm i\text{PV}\frac{1}{\epsilon},
\end{align}
where $\text{PV}$ means the principal value. Hence,
\begin{gather}\label{c_Heisenberg_equation_solution_approx2}
\begin{aligned}
&c_\omega 
=c_\omega(0)e^{-i\omega t}
- i\sum_{m=1}^{N} \sum_{j,j'}g_m(\omega)\\
&\times \Bigg(C_{m,jj'}\sigma_{m,jj'}\bigg(\delta(\omega-\omega_{m,j'j})-i\text{PV}\frac{1}{\omega-\omega_{m,j'j}}\bigg)\\
&-C_{m,jj'}^*\sigma_{m,j'j}\bigg(\delta(\omega+\omega_{m,j'j})-i\text{PV}\frac{1}{\omega+\omega_{m,j'j}}\bigg)\Bigg).
\end{aligned}
\end{gather}
When inserting the above into Eq.~\eqref{Q_equation}, two different integrals over the 
frequency need to be performed,
\begin{subequations}
\begin{align}
&\begin{aligned}&\text{Re}[\xi_{mn,j'j}]=2\pi\int_0^\infty g_m(\omega)g_n(\omega)\\
&\times(\delta(\omega-\omega_{n,j'j})-\delta(\omega+\omega_{n,j'j}))\dif \omega,
\end{aligned}\displaybreak[0]\\
&\text{Im}[\xi_{mn,j'j}]=-2\text{PV}\int_0^\infty g_m(\omega)g_n(\omega)\frac{2\omega}{\omega^2-\omega_{n,j'j}^2}\dif\omega;
\end{align}
\label{Re_xi_Im_xi_integrals}
\end{subequations}
and then the same Eq.~\eqref{Q_equation3} is obtained with 
$\mathcal{O}_{mn}=\sum_{j,j'}\xi_{mn,j'j}C_{n,jj'}\sigma_{n,jj'}$, but 
$\xi_{mn,j'j}$ is given by the integrals~\eqref{Re_xi_Im_xi_integrals} 
instead of Eq.~\eqref{xi_def}.
The integrands of both of the integrals~\eqref{Re_xi_Im_xi_integrals} are different if
the rotating wave approximation is performed already in the 
Hamiltonian~\eqref{H_i}, again illustrating the importance of delaying this 
approximation until after these integrals are evaluated~\cite{ott_pra13}. 

To evaluate the integrals~\eqref{Re_xi_Im_xi_integrals}, a particular form 
of the coupling $g_m(\omega)$ needs to be chosen. Using 
$g_m(\omega)=G_m(\sqrt{\omega}/\sqrt{1+\mathcal{A}\omega^2})\cos(k_\omega x_m)$, 
we get the real part
\begin{align}
&\begin{aligned}
&\text{Re}[\xi_{mn,j'j}]=\pi G_m G_n
\frac{\omega_{n,j'j}}{1+\mathcal{A}\omega_{n,j'j}^2}\\
&\times\left(\cos(k_{|\omega_{n,j'j}|} (x_m-x_n))+\cos(k_{|\omega_{n,j'j}|} (x_m-x_n))\right).
\end{aligned}
\end{align}
For the imaginary part, we first switch to the integration over 
$k=\omega/v_\text{g}$, so that we need to evaluate the integral of the form
\begin{align}\label{I_integral1}
I(x)=\text{PV}\int_0^\infty 
\frac{2k^2}{(1+\mathcal{A}v_\text{g}^2k^2)(k^2-k_{n,j'j}^2)}
\cos(k x)\dif k,
\end{align}
where $k_{n,j'j}=k_{\omega_{n,j'j}}$. Then 
\begin{gather}
\text{Im}[\xi_{mn,j'j}]=-G_m G_n v_\text{g}\left(I(x_m-x_n)+I(x_m+x_n)\right).
\end{gather}

\begin{figure}[t]
\begin{center}
\includegraphics{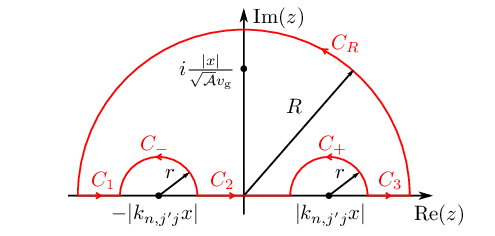}
\end{center}
\caption{The complex contour (red) used for evaluation of the 
integral~\eqref{I_integral2}. The 3 relevant poles of the 
integrand~\eqref{f_I_integrand} at $-|k_{n,j'j}x|$, $|k_{n,j'j}x|$, and 
$i|x|/(\sqrt{\mathcal{A}}v_\text{g})$ are shown as dots. The half-circle $C_R$ 
is parametrized by the radius $R\rightarrow\infty$. The half-circles $C_\pm$ 
are parametrized by the radius $r\rightarrow 0$. The direction of integration along the sections is shown with arrows.
\label{fig_complex_contour}}
\end{figure}

The evaluation of the integral~\eqref{I_integral1} can be done by integrating 
in the complex plane and using the residue theorem.  After further variable 
changes, the integral is written
\begin{gather}\label{I_integral2}
I(x)=\frac{1}{|x|}\text{PV}\int_{-\infty}^\infty f_I(z)\dif z,
\end{gather}
with the integrand
\begin{gather}\label{f_I_integrand}
f_I(z)=\frac{2(\frac{z}{x})^2}
{(1+\mathcal{A}v_\text{g}^2(\frac{z}{x})^2)
((\frac{z}{x})^2-k_{n,j'j}^2)}
e^{iz}.
\end{gather}
The integral~\eqref{I_integral2} is then evaluated using the complex contour 
shown in Fig.~\ref{fig_complex_contour}. The principal value is found by 
using half-circles with a radius $r$ around the poles of $f_I$ lying 
on the real line and then letting $r\rightarrow 0$. A large half-circle with 
the radius $R\rightarrow\infty$ encloses the pole on the positive imaginary axis.

Defining
\begin{gather}
I_C=\lim_{R\rightarrow\infty, r\rightarrow 0}\frac{1}{|x|}\text{PV}\int_C f_I(z)\dif z
\end{gather} 
to be the integral along each section $C$ of the contour, and $\text{Res}(f_I(z),c)$ 
to be the residue of the function $f_I$ at $z=c$, we see that the integral to 
do is
\begin{gather}
I(x)=I_{C_1}+I_{C_2}+I_{C_3},
\end{gather}
while for the entire contour, it holds that
\begin{gather}
\begin{aligned}
&I_{C_1}-I_{C_-}+I_{C_2}-I_{C_+}+I_{C_3}+I_{C_R}\\
&=\frac{2\pi i}{|x|}\text{Res}\left(f_I(z),
i\frac{|x|}{\sqrt{\mathcal{A}}v_\text{g}}\right),
\end{aligned}
\end{gather}
where the signs are determined according to the directions shown by the arrows 
in Fig.~\ref{fig_complex_contour}.

For large $|z|$, it holds that $|f_I(x)|\leq M/|z|^2$ with 
$M=2x^2/(\mathcal{A}v_\text{g}^2)$. This is a sufficient condition to show 
that $I_{C_R}=0$. Therefore, we have
\begin{gather}
I(x)=\frac{2\pi i}{|x|}\text{Res}\left(f_I(z),
i\frac{|x|}{\sqrt{\mathcal{A}}v_\text{g}}\right)
+I_{C_+}+I_{C_-},
\end{gather}
where
\begin{gather}
I_{C_\pm} = \frac{\pi i}{|x|}\text{Res}\left(f_I(z),
\pm|k_{n,j'j}x|\right).
\end{gather}
For simple poles, like we have here, it holds that 
$\text{Res}(f_I(z),c)=\lim_{z\rightarrow c} (z-c)f_I(z)$. Hence,
\begin{gather}
\begin{aligned}
I(x)=\frac{\pi}{1+\mathcal{A}\omega_{n,j'j}^2}
\Bigg(&\frac{e^{-|x|/(\sqrt{\mathcal{A}}v_\text{g})}}{\sqrt{\mathcal{A}}v_\text{g}}\\
&-k_{n,j'j}\sin(k_{n,j'j}|x|)\Bigg),
\end{aligned}
\end{gather}
and
\begin{gather}\label{xi_def_A}
\begin{aligned}
&\xi_{mn,j'j}=\frac{\pi G_m G_n }{1+\mathcal{A}\omega_{n,j'j}^2}\\
&\times\Bigg(
-\frac{ie^{-|x_m-x_n|/\left(\sqrt{\mathcal{A}}v_\text{g}\right)}}{\sqrt{\mathcal{A}}}
+\omega_{n,j'j}e^{ik_{n,j'j} |x_m-x_n|}\\
&-\frac{ie^{-|x_m-x_n|/\left(\sqrt{\mathcal{A}}v_\text{g}\right)}}{\sqrt{\mathcal{A}}}
+\omega_{n,j'j}e^{ik_{n,j'j} |x_m+x_n|}\Bigg).
\end{aligned}
\end{gather}
Therefore, for
$g_m(\omega)=G_m(\sqrt{\omega}/\sqrt{1+\mathcal{A}\omega^2})\cos(k_\omega x_m)$ 
with $\mathcal{A}>0$ and following the same steps as in App.~\ref{App:master_equation} after 
Eq.~\eqref{Q_equation3}, we see that the factor $\xi_{mn,j'j}$ in the master 
equation~\eqref{master_equation} is given by Eq.~\eqref{xi_def_A} instead of 
Eq.~\eqref{xi_def}. For $A\rightarrow 0^+$ and 
$G_m=\sqrt{\Gamma_m/(2\pi\omega_m)}$, Eq.~\eqref{xi_def_A} becomes 
Eq.~\eqref{xi_def}, as expected.

\section{Delay differential equations for the decay}\label{App:dde_decay}
In this appendix, we derive the model for the decay without the
approximations~\eqref{free_evolution_appr}~or~\eqref{driven_evolution_appr},
following the approach of Ref.~\cite{koshino_prapplied20}. In the 
single-excitation subspace, only 3 subsystem states are relevant: a single 
excitation in either of the transmons or the resonator. Using the diagonal 
basis, the subsystem $1$ consisting of the transmon and the resonator is represented by the two 
eigenstates~\eqref{H_sm_aa_res_single_excitation_eigvecs} with the 
eigenfrequencies~\eqref{H_sm_aa_res_single_excitation_eigvals}. For the subsystem $2$ (JQF), there 
is only one eigenstate~$|1_2\rangle$ to consider. 
In the single-excitation subspace, we therefore have 
\begin{subequations}
\begin{align}
&\mathcal{O}_1=a=C_{1,01}\sigma_{1,01}+C_{1,02}\sigma_{1,02},\displaybreak[0]\\
&\mathcal{O}_2=b_2=C_{2,01}\sigma_{2,01},
\end{align}
\label{O_operators_single_excitation}
\end{subequations}
where $C_{1,01}=\langle 0_1|a|1_1\rangle=\sin(\theta)$, 
$C_{1,02}=\langle 0_1|a|2_1\rangle=\cos(\theta)$, and 
$C_{2,01}=\langle 0_2|b_2|1_2\rangle=1$.

Equations of motion for the operators $\sigma_{m,0j}$ are obtained from 
Eq.~\eqref{Q_equation2} under the approximation~\eqref{dot_O_approx}. We 
consider the single-excitation state
\begin{gather}
\begin{aligned}
&|\psi(t)\rangle=\sum_{m,j} \alpha_{m,j}(t)\sigma_{m,j0}(0)|\text{vac}\rangle\\
&+\int_0^\infty f_\omega(t) c_\omega^\dagger(0)\dif\omega|\text{vac}\rangle,
\end{aligned}
\end{gather}
where $|\text{vac}\rangle=|0_1\rangle|0_2\rangle|\text{vac}_c\rangle$, and $|\text{vac}_c\rangle$ 
is the vacuum state for the transmission line. The equations of motion for 
the amplitudes $\alpha_{m,j}$ are found as 
\begin{gather}
\dot{\alpha}_{m,j}(t)=\langle\text{vac}|\dot{\sigma}_{m,0j}(t)|\psi(0)\rangle.
\end{gather}
Defining the inputs
$f_{\text{in},m}(t)=\int_0^\infty g_m(\omega) f_\omega(0)e^{-i\omega t}\dif \omega$,
slowly-varying quantities
$\alpha_{m,j}(t)=\tilde{\alpha}_{m,j}(t)e^{-i\omega_{1,10}t}$ and 
$f_{\text{in},m}(t)=\tilde{f}_{\text{in},m}(t)e^{-i\omega_{1,10}t}$, and setting 
$x_1=0$ for simplicity, gives the delay differential equations
\begin{widetext}
\begin{subequations}
\begin{align}
&\begin{aligned}
&\dot{\tilde{\alpha}}_{1,1}(t)
=-iC_{1,01}^*\tilde{f}_{\text{in},1}(t)
-|C_{1,01}|^2\frac{\kappa}{2}\frac{\omega_{1,10}}{\omega_\text{r}}\tilde{\alpha}_{1,1}(t)
-C_{1,01}^*C_{1,02}\frac{\kappa}{2}\frac{\omega_{1,20}}{\omega_\text{r}}\tilde{\alpha}_{1,2}(t)\\
&-C_{1,01}^*C_{2,01}\frac{\sqrt{\kappa\gamma_2}}{2}
\frac{\omega_{2,10}}{\sqrt{\omega_\text{r}\omega_{\text{t},2}}}e^{ik_{\omega_{1,10}}x_2}
\tilde{\alpha}_{2,1}(t-x_2/v_\text{g}),
\end{aligned}\displaybreak[0]\\
&\begin{aligned}
&\dot{\tilde{\alpha}}_{1,2}(t)
=-i(\omega_{1,20}-\omega_{1,10})\tilde{\alpha}_{1,2}(t)
-iC_{1,02}^*\tilde{f}_{\text{in},1}(t)
-|C_{1,02}|^2\frac{\kappa}{2}\frac{\omega_{1,20}}{\omega_\text{r}}\tilde{\alpha}_{1,2}(t)
-C_{1,02}^*C_{1,01}\frac{\kappa}{2}\frac{\omega_{1,10}}{\omega_\text{r}}\tilde{\alpha}_{1,1}(t)\\
&-C_{1,02}^*C_{2,01}\frac{\sqrt{\kappa\gamma_2}}{2}
\frac{\omega_{2,10}}{\sqrt{\omega_\text{r}\omega_{\text{t},2}}}e^{ik_{\omega_{1,10}}x_2}
\tilde{\alpha}_{2,1}(t-x_2/v_\text{g}),
\end{aligned}\displaybreak[0]\\
&\begin{aligned}
&\dot{\tilde{\alpha}}_{2,1}(t)
=-i(\omega_{2,10}-\omega_{1,10})\tilde{\alpha}_{2,1}(t)
-iC_{2,01}^*\tilde{f}_{\text{in},2}(t)
-|C_{2,01}|^2\frac{\gamma_2}{4}\frac{\omega_{2,10}}{\omega_{\text{t},2}}\left(\tilde{\alpha}_{2,1}(t)+e^{2ik_{\omega_{1,10}}x_2}\tilde{\alpha}_{2,1}(t-2x_2/v_\text{g})\right)\\
&-C_{2,01}^*C_{1,01}\frac{\sqrt{\kappa\gamma_2}}{2}\frac{\omega_{1,10}}{\sqrt{\omega_\text{r}\omega_{\text{t},2}}}e^{ik_{\omega_{1,10}}x_2}\tilde{\alpha}_{1,1}(t-x_2/v_\text{g})
-C_{2,01}^*C_{1,02}\frac{\sqrt{\kappa\gamma_2}}{2}\frac{\omega_{2,10}}{\sqrt{\omega_\text{r}\omega_{\text{t},2}}}e^{ik_{\omega_{1,10}}x_2}\tilde{\alpha}_{1,2}(t-x_2/v_\text{g}),
\end{aligned}
\end{align}
\label{decay_dde}
\end{subequations}
\end{widetext}
where the appropriate Heaviside theta function factors on the delayed terms are implicit. 

\begin{figure}[t]
\begin{center}
\includegraphics{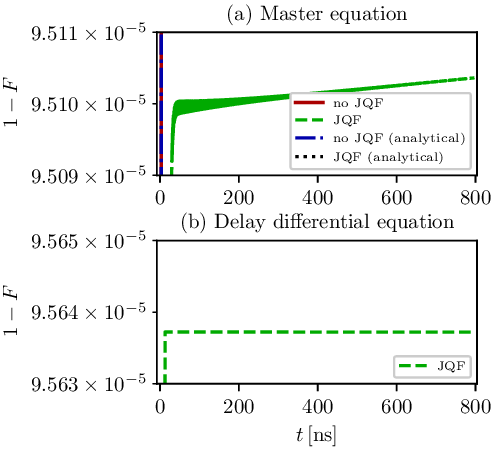}
\end{center}
\caption{(a) The zoomed-in version of Fig.~\ref{fig_decay}, using the master 
equation~\eqref{master_equation} for the numerical results. Rapid oscillations 
of the dashed green curve make it appear thick for smaller times. The horizontal 
dotted black line showing Eq.~\eqref{F_dark} is 
below 
the lower limit of the vertical axis by about ${2\times 10^{-8}}$ and hence cannot be seen. (b) The setup with 
a JQF corresponding to the dashed green curve of (a), but calculated using the 
delay differential equations~\eqref{decay_dde}. In (b), $F=|\tilde{\alpha}_{1,1}(t)|^2$.
\label{fig_dde}}
\end{figure}

We set the initial conditions $\tilde{\alpha}_{1,1}(0)=1$, 
$\tilde{\alpha}_{1,2}(0)=\tilde{\alpha}_{2,1}(0)=0$, $f_\omega(0)=0$, and use the Euler method 
for the numerical solution. The Runge-Kutta method applied to the delay 
differential equations requires accurate interpolations between the time 
steps~\cite{oberle_nm81}. For the Euler method, if the delays $x_2/v_\text{g}$ 
are an integer multiple of the step size, no such interpolation is required. 
The drawback is a significantly larger number of time steps $N_t$ required to 
reach convergence. We use $N_t=4\times 10^{12}$ in Fig.~\ref{fig_dde}(b). Compared to 
the master equation curves in Fig.~\ref{fig_dde}(a), the ``steady-state'' 
value in Fig.~\ref{fig_dde}(b) is shifted by about $5\times 10^{-7}$.

\section{Derivation of the reflection coefficient}\label{App:reflection_coefficient}
In this appendix, we derive the input-output relations corresponding to the master 
equation~\eqref{master_equation}. As in App.~\ref{App:master_equation}, the 
general setup with $N$ attached subsystems is considered, generalizing from 
the case $N=2$ in the main text. The voltage operator in 
the transmission line is
\begin{gather}
V(x)=-i\sqrt{\frac{\hbar Z_0}{\pi}}\int_0^\infty 
\sqrt{\omega}(c_\omega-c_\omega^\dagger)\cos(k_\omega x)\dif \omega,
\end{gather}
where $Z_0$ is the impedance of the transmission line. With the 
charge number operators 
$n_m\propto -i(\mathcal{O}_m-\mathcal{O}_m^\dagger)$ for the
subsystems attached to the transmission line, the above expression for $V(x)$ determines the interaction Hamiltonian 
$H_\text{i}\propto \sum_{m=1}^N V(x_m)n_m$, resulting 
in Eq.~\eqref{H_i}. This can be shown by performing the circuit quantization 
of the setup (ignoring the $A^2$~term). 

The voltage operator is split into the 
right-moving~($V_+$) and left-moving~($V_-$) parts, $V(x)=V_+(x)+V_-(x)$, 
where
\begin{gather}
V_\pm(x)=-i\sqrt{\frac{\hbar Z_0}{4\pi}}\int_0^\infty 
\sqrt{\omega}(c_\omega e^{\pm ik_\omega x}-c_\omega^\dagger e^{\mp ik_\omega x})\dif \omega.
\end{gather}
We have
\begin{gather}\label{c_plus_c_dagger_V_plus}
\begin{aligned}
&c_\omega e^{\pm ik_\omega x}-c_\omega^\dagger e^{\mp ik_\omega x}\\
&= c_\omega(0)e^{-i\omega t\pm ik_\omega x}-c_\omega^\dagger(0)e^{i\omega t\mp ik_\omega x}
-i\sum_{m=1}^N g_m(\omega)\\
&\times\Bigg(\int_0^t\mathcal{O}_m(t-t')
\left(e^{-i\omega t'\pm ik_\omega x}-e^{i\omega t'\mp ik_\omega x}\right)\dif t'\\
&-\int_0^t\mathcal{O}_m^\dagger(t-t')
\left(e^{-i\omega t'\pm ik_\omega x}-e^{i\omega t'\mp ik_\omega x}\right)\dif t'\Bigg).
\end{aligned}
\end{gather}
The same comment about Eq.~\eqref{c_plus_c_dagger} 
applies to Eq.~\eqref{c_plus_c_dagger_V_plus}---that additional terms are 
present due to not making the rotating wave approximation in the 
Hamiltonian~\eqref{H_i}. 

Only after calculating 
$c_\omega e^{\pm ik_\omega x}-c_\omega^\dagger e^{\mp ik_\omega x}$, the 
resulting expression can be split into two parts involving either creation or
annihilation operators. I.e., we write 
$V_\pm(x)=\mathcal{V}_\pm(x)+\mathcal{V}_\pm^\dagger(x)$, where
\begin{gather}\label{mathcal_V_pm}
\begin{aligned}
&\mathcal{V}_\pm(x)=\mathcal{V}_{\pm,0}(x)\\
&-\sqrt{\frac{\hbar Z_0}{4\pi}}\sum_{m=1}^N \int_0^t\mathcal{O}_m(t-t')\int_0^\infty\sqrt{\omega}g_m(\omega)\\
&\times
\left(e^{-i\omega t'\pm ik_\omega x}-e^{i\omega t'\mp ik_\omega x}\right)\dif \omega\dif t',
\end{aligned}
\end{gather}
and 
\begin{gather}
\mathcal{V}_{\pm,0}(x)=-i\sqrt{\frac{\hbar Z_0}{4\pi}}\int_0^\infty 
\sqrt{\omega}c_\omega(0) e^{-i\omega t \pm ik_\omega x}\dif \omega.
\end{gather}
The integral over $\omega$ in Eq.~\eqref{mathcal_V_pm} is 
similar to the integral in Eq.~\eqref{delta_prime_integral}, and we get
\begin{gather}\label{delta_prime_integral_V_pm}
\begin{aligned}
&\int_0^\infty \sqrt{\omega}g_m(\omega)\left(e^{-i\omega t'\pm ik_\omega x}
-e^{i\omega t'\mp ik_\omega x}\right)\dif \omega\\
&=i\pi G_m\Bigg(
\dot{\delta}\bigg(t'\mp\frac{x-x_m}{v_\text{g}}\bigg)
+\dot{\delta}\bigg(t'\mp\frac{x+x_m}{v_\text{g}}\bigg)\Bigg).
\end{aligned}
\end{gather}

While up to now the calculation for the right-moving and left-moving parts was 
symmetric, the asymmetry arises after the integration over $t'$. Using 
Eq.~\eqref{delta_dot_property},
\begin{subequations}
\begin{align}
&\begin{aligned}
&\mathcal{V}_+(x)=\mathcal{V}_{+,0}(x)-\sqrt{\frac{\hbar Z_0}{4\pi}}\sum_{m=1}^N
(i\pi) G_m\\
&\times
\Bigg(\dot{\mathcal{O}}_m\bigg(t-\frac{x-x_m}{v_\text{g}}\bigg)
\theta_\text{H}\bigg(t-\frac{x-x_m}{v_\text{g}}\bigg)
\theta_\text{H}(x-x_m)\\
&+\dot{\mathcal{O}}_m\bigg(t-\frac{x+x_m}{v_\text{g}}\bigg)
\theta_\text{H}\bigg(t-\frac{x+x_m}{v_\text{g}}\bigg)\Bigg),
\end{aligned}\displaybreak[0]\\
&\begin{aligned}
&\mathcal{V}_-(x)=\mathcal{V}_{-,0}(x)-\sqrt{\frac{\hbar Z_0}{4\pi}}\sum_{m=1}^N
(i\pi) G_m \\
&\times
\dot{\mathcal{O}}_m\bigg(t-\frac{x_m-x}{v_\text{g}}\bigg)
\theta_\text{H}\bigg(t-\frac{x_m-x}{v_\text{g}}\bigg)
\theta_\text{H}(x_m-x).
\end{aligned}
\end{align}
\label{mathcal_V_pm_2}
\end{subequations}
The right-moving 
part $\mathcal{V}_+$ has two types of contributions: those that are emitted directly to 
the right and those that are emitted to the left and then reflected from the 
boundary at $x=0$. The left-moving part $\mathcal{V}_-$ only has
contributions from the emission directly to the left.

Under the approximations~\eqref{dot_O_approx}~and~\eqref{driven_evolution_appr},
\begin{subequations}
\begin{align}
&\begin{aligned}
&\mathcal{V}_+(x)=\mathcal{V}_{+,0}(x)-\sqrt{\frac{\hbar Z_0}{4\pi}}\sum_{m=1}^N\sum_{j,j'}
\pi \omega_{m,j'j} G_m C_{m,jj'}\\
&\times\sigma_{m,jj'}
\left(\theta_\text{H}(x-x_m)e^{ik_{\omega_\text{d}}(x-x_m)}+e^{ik_{\omega_\text{d}}(x+x_m)}\right),
\end{aligned}\displaybreak[0]\\
&\begin{aligned}
&\mathcal{V}_-(x)=\mathcal{V}_{-,0}(x)-\sqrt{\frac{\hbar Z_0}{4\pi}}\sum_{m=1}^N\sum_{j,j'}
\pi \omega_{m,j'j} G_m C_{m,jj'}\\
&\times\sigma_{m,jj'}
\theta_\text{H}(x_m-x)e^{ik_{\omega_\text{d}}(x_m-x)}.
\end{aligned}
\end{align}
\label{mathcal_V_pm_3}
\end{subequations}

The reflection coefficient is defined to be
\begin{gather}
r=\frac{\tr[\mathcal{V}_+(x_{N}^+)\rho]}{\tr[\mathcal{V}_-(x_{N}^+)\rho]},
\end{gather}
where $x_{N}^+=x_{N}+\epsilon$ with $\epsilon>0$ such that $\epsilon\rightarrow 0$ at the end of the 
calculation. We assume that the positions~$x_m$ are ordered such that they 
increase with increasing $m$, and hence $x_{N}^+$ is the position just to the 
right of the last subsystem attached to the transmission line. Hence,
$\mathcal{V}_-(x_{N}^+)=\mathcal{V}_{-,0}(x_{N}^+)$. Using 
Eq.~\eqref{c_omega_action_on_ket_alpha}, noting that due to 
Eq.~\eqref{H_0_condition}, we have
$\tr[\sigma_{m,jj'}\rho]
=\tr_\text{s}[\sigma_{m,jj'}\tilde{\rho}_\text{s}]e^{-i\omega_\text{d}t}$,
and removing the overall propagation phase $e^{2ik_{\omega_\text{d}}x_{N}^+}$,
the reflection coefficient
\begin{gather}\label{reflection_coefficient_flux}
\begin{aligned}
&r=1
-i\sum_{m=1}^N\sum_{j,j'}
\frac{\omega_{m,j'j}}{\sqrt{\omega_\text{d}\omega_m}}\sqrt{\frac{\Gamma_m}{\dot{n}}}
C_{m,jj'}\\
&\times \tr_\text{s}[\sigma_{m,jj'}\tilde{\rho}_\text{s}]
\cos(k_{\omega_\text{d}}x_m)e^{-i\phi}
\end{aligned}
\end{gather}
is obtained. Writing the photon flux $\dot{n}$ in terms of the reference Rabi frequency 
$\Omega_1$ using Eq.~\eqref{Omega_m_def} with $x_1=0$, results in the
expression~\eqref{reflection_coefficient} of the main text.

\section{Calculation of the gradient}\label{App:optimal_control}
In this appendix, we give details about the calculation of the gradient for 
the optimal control approach used in Sec.~\ref{sec_control} of the main text. 
The master equation~\eqref{master_equation} is rewritten such that the 
elements of the density matrix $\tilde{\rho}_\text{s}$ are arranged as a 
vector $\vec{\rho}_\text{s}$, resulting in
$\dot{\vec{\rho}}_\text{s}=L(t)\vec{\rho}_\text{s}$. In the same way, the 
matrices $M_{\tilde{F}}=UAU^\dagger$ and $\rho_\text{s}(t)=\mathcal{M}(A)$ in Eq.~\eqref{F_tilde_def} are also 
written as 
vectors $\vec{M}_{\tilde{F}}$ and $\vec{\rho}_\text{s}(t)$, and hence we can write Eq.~\eqref{F_tilde_def} 
as the inner product
\begin{gather}\label{F_tilde_def_vec}
\tilde{F}=\vec{M}_{\tilde{F}}^\dagger \vec{\rho}_\text{s}(t),
\end{gather}
where we have used the fact that $M_{\tilde{F}}$ is a Hermitian matrix. The 
latter follows from $A$ either being a density matrix or one of the Pauli 
matrices, and $U$ being unitary.

We solve the master equation with the 4$^\text{th}$ order Runge-Kutta method and 
use the reverse mode automatic differentiation to calculate the gradient of 
Eq.~\eqref{F_tilde_def_vec} with respect to the Fourier amplitudes $a_p$ and 
$b_p$ in the pulse shape parametrization~\eqref{Re_Im_Omega_parametrization}. The gradient of the average fidelity~\eqref{F_tilde_average_def} is 
then obtained by adding the contributions for all the initial matrices. Compared 
to the general case of the nonlinear differential equations solved by the 
Runge-Kutta method~\cite{evtushenko_oms98,walther_coa07}, the linearity of the 
master equation allows for an approach that is superficially similar to the 
other optimal control approaches where forward propagation of the equations of motion is alternated with 
backward propagation of the adjoint equations~\cite{kosloff_chemphys89,somloi_chemphys93}. In 
contrast to Refs.~\cite{kosloff_chemphys89,somloi_chemphys93}, however, the derivation of the 
adjoint equation is more involved than simply taking the adjoint of $L(t)$.

Defining $\Delta t=t_\text{f}/N_t$,  
$t_n=(\Delta t)n$, $\vec{\rho}_n=\vec{\rho}_\text{s}(t_n)$, $L_{1,n}=L(t_n)\Delta t$, 
$L_{2,n}=L(t_n+(\Delta t)/2)\Delta t$, $L_{3,n}=L(t_{n+1})\Delta t$, the 4$^\text{th}$ order Runge-Kutta method can be written
\begin{subequations}
\begin{align}
&\vec{k}_{1,n}=L_{1,n}\vec{\rho}_n,\\
&\vec{k}_{2,n}=L_{2,n}(\vec{\rho}_n+\vec{k}_{1,n}/2),\\
&\vec{k}_{3,n}=L_{2,n}(\vec{\rho}_n+\vec{k}_{2,n}/2),\\
&\vec{k}_{4,n}=L_{3,n}(\vec{\rho}_n+\vec{k}_{3,n}),\\
&\vec{\rho}_{n+1}=\vec{\rho}_{n}+\vec{k}_{1,n}/6+\vec{k}_{2,n}/3+\vec{k}_{3,n}/3+\vec{k}_{4,n}/6.
\end{align}
\label{runge_kutta_4}
\end{subequations}
The reverse mode automatic differentiation applied to this algorithm requires storing or recomputing (during 
the backward propagation) the vectors $\vec{\rho}_{n}$, $\vec{k}_{1,n}$, 
$\vec{k}_{2,n}$, $\vec{k}_{3,n}$, and $\vec{k}_{4,n}$ for all $n$. The vectors 
$\vec{\rho}_{n}$ can be recomputed after the forward propagation by applying 
the Runge-Kutta method backward in time, starting from $\vec{\rho}_{N_t}$. In 
our numerical simulations, we store as many of the vectors $\vec{\rho}_{n}$ as 
could be fit into memory, uniformly spaced over all the time indices $n$. The 
vectors $\vec{\rho}_{n}$ between the stored ones are recomputed by applying the 
Runge-Kutta method backward in time. We find that this decreases the 
numerical error due to inexact recomputation of the vectors $\vec{\rho}_{n}$.

The vectors $\vec{k}_{1,n}$, $\vec{k}_{2,n}$, $\vec{k}_{3,n}$, and 
$\vec{k}_{4,n}$ are always recomputed but in an indirect way. We rewrite 
Eqs.~\eqref{runge_kutta_4} into the form
\begin{gather}
\vec{\rho}_{n+1}=K_n\vec{\rho}_n, 
\end{gather}
where
\begin{gather}
\begin{aligned}
&K_n = I+\frac{1}{6}L_{1,n}+\frac{1}{3}\del{L_{2,n}+\frac{1}{2}L_{2,n}L_{1,n}}\\
&+\frac{1}{3}\del{L_{2,n}+\frac{1}{2}L_{2,n}L_{2,n}+\frac{1}{4}L_{2,n}L_{2,n}L_{1,n}}\\
&+\frac{1}{6}\left(L_{3,n}+L_{3,n}L_{2,n}+\frac{1}{2}L_{3,n}L_{2,n}L_{2,n}\right.\\
&\left.+\frac{1}{4}L_{3,n}L_{2,n}L_{2,n}L_{1,n}\right),
\end{aligned}
\end{gather}
and apply the reverse mode automatic differentiation on this form.

The gradient of Eq.~\eqref{F_tilde_def_vec} at the final time $t=t_\text{f}$ is 
\begin{gather}
\begin{aligned}
\dpd{\tilde{F}}{a_p}=\vec{M}_{\tilde{F}}^\dagger\dpd{\vec{\rho}_{N_t}}{a_p},&&
\dpd{\tilde{F}}{b_p}=\vec{M}_{\tilde{F}}^\dagger\dpd{\vec{\rho}_{N_t}}{b_p},
\end{aligned}
\end{gather}
where
\begin{subequations}
\begin{align}
&\dpd{\vec{\rho}_{n}}{a_p}
=\dpd{K_{n-1}}{a_p}\vec{\rho}_{n-1}
+K_{n-1}\dpd{\vec{\rho}_{n-1}}{a_p},\\
&\dpd{\vec{\rho}_{n}}{b_p}
=\dpd{K_{n-1}}{b_p}\vec{\rho}_{n-1}
+K_{n-1}\dpd{\vec{\rho}_{n-1}}{b_p}.
\end{align}
\end{subequations}
By substituting these equations into themselves for all $n$ and defining
the initial value $\vec{\chi}_{N_t}^\dagger=\vec{M}_{\tilde{F}}^\dagger$ and the adjoint equation
\begin{gather}\label{adjoint_equation}
\vec{\chi}_{n-1}^\dagger=\vec{\chi}_{n}^\dagger K_{n-1},
\end{gather}
we end up with
\begin{subequations}
\begin{align}
&\dpd{\tilde{F}}{a_p}
=\sum_{n=1}^{N_t}\vec{\chi}_n^\dagger\dpd{K_{n-1}}{a_p}\vec{\rho}_{n-1},\\
&\dpd{\tilde{F}}{b_p}
=\sum_{n=1}^{N_t}\vec{\chi}_n^\dagger\dpd{K_{n-1}}{b_p}\vec{\rho}_{n-1},
\end{align}
\label{grad_F_def_sum}
\end{subequations}
where the sums are can be efficiently evaluated by starting with $n=N_t$ and 
propagating $\vec{\chi}_{n}$ backward using Eq.~\eqref{adjoint_equation}. 
We give more details below, but first we summarize the entire procedure:
\begin{enumerate}
\item Propagate forward using Eqs.~\eqref{runge_kutta_4}, saving 
as many of the intermediate values $\vec{\rho}_n$, as can be fit into memory.
\item Initialize $n=N_t$, and use $\vec{\chi}_{N_t}^\dagger=\vec{M}_{\tilde{F}}^\dagger$.
\item If $\vec{\rho}_{n-1}$ is not stored in memory, calculate it by 
propagating Eqs.~\eqref{runge_kutta_4} backward in time, otherwise use the stored $\vec{\rho}_{n-1}$.
\item Evaluate the scalars given by Eqs.~\eqref{scalars_S_def}.
\item Add the contributions from this $n$ to the gradient using 
Eqs.~\eqref{grad_F_def_single_term} for all $a_p$ and $b_p$.
\item Calculate $\vec{\chi}_{n-1}$ using 
Eq.~\eqref{adjoint_equation_expanded}.
\item If $n>0$, go to step 3 replacing $n$ with $n-1$. Otherwise, stop.
\end{enumerate}
The above procedure needs a constant number of the computationally expensive matrix-vector 
multiplications for every time index $n$, 
independent of the number of the parameters $a_p$ and $b_p$. 

To derive the expressions for the above procedure, we first note that
\begin{subequations}
\begin{align}
&(\Delta t)\dpd{L(t)}{a_p} = \dpd{\text{Re}[\Omega]}{a_p}(t)T_\text{Re},\displaybreak[0]\\
&(\Delta t)\dpd{L(t)}{b_p} = \dpd{\text{Im}[\Omega]}{b_p}(t)T_\text{Im},
\end{align}
\end{subequations}
where, assuming that the density matrix is written as a vector in the
row-major form (i.e., $(\tilde{\rho}_\text{s})_{l,l'}=(\vec{\rho}_\text{s})_{lN_\text{b}+l'}$,
$N_\text{b}$ is the Hilbert space basis size, and $0\leq l,l'\leq N_\text{b}-1$),
\begin{subequations}
\begin{align}
&T_\text{Re}=-\frac{i}{\hbar}(\Delta t)(\tilde{H}_\text{d,Re}\otimes I - I\otimes \tilde{H}_\text{d,Re}^T),\displaybreak[0]\\
&T_\text{Im}=-\frac{i}{\hbar}(\Delta t)(\tilde{H}_\text{d,Im}\otimes I - I\otimes \tilde{H}_\text{d,Im}^T),
\end{align}
\end{subequations}
and $\tilde{H}_\text{d,Re}$, $\tilde{H}_\text{d,Im}$ are given by Eqs.~\eqref{H_d_Re_Im_def}.
We define the following temporary vectors
\begin{subequations}
\begin{align}
&\vec{l}_{0,n}=T_\text{Re}\vec{\rho}_{n},
&&\vec{l}_{1,n}=L_{1,n}\vec{\rho}_{n},\displaybreak[0]\\
&\vec{l}_{2,n}=L_{2,n}\vec{\rho}_{n},
&&\vec{l}_{3,n}=T_\text{Re}\vec{l}_{1,n},\displaybreak[0]\\
&\vec{l}_{4,n}=T_\text{Re}\vec{l}_{2,n},
&&\vec{l}_{5,n}=L_{2,n}\vec{l}_{0,n},\displaybreak[0]\\
&\vec{l}_{6,n}=L_{2,n}\vec{l}_{1,n},
&&\vec{l}_{7,n}=L_{2,n}\vec{l}_{3,n},\displaybreak[0]\\
&\vec{l}_{8,n}=L_{2,n}\vec{l}_{5,n},
&&\vec{l}_{9,n}=T_\text{Re}\vec{l}_{6,n};\displaybreak[0]\\
&\vec{m}_{0,n}=T_\text{Im}\vec{\rho}_{n},
&&\vec{m}_{1,n}=L_{1,n}\vec{\rho}_{n},\displaybreak[0]\\
&\vec{m}_{2,n}=L_{2,n}\vec{\rho}_{n},
&&\vec{m}_{3,n}=T_\text{Im}\vec{m}_{1,n},\displaybreak[0]\\
&\vec{m}_{4,n}=T_\text{Im}\vec{m}_{2,n},
&&\vec{m}_{5,n}=L_{2,n}\vec{m}_{0,n},\displaybreak[0]\\
&\vec{m}_{6,n}=L_{2,n}\vec{m}_{1,n},
&&\vec{m}_{7,n}=L_{2,n}\vec{m}_{3,n},\displaybreak[0]\\
&\vec{m}_{8,n}=L_{2,n}\vec{m}_{5,n},
&&\vec{m}_{9,n}=T_\text{Im}\vec{m}_{6,n};
\end{align}
\end{subequations}
and scalars
\begin{widetext}
\begin{subequations}
\begin{align}
&\begin{aligned}
&S_{1,n,\text{Re}}
=\vec{\chi}_n^\dagger\left(
\frac{1}{6}\vec{l}_{0,n-1}
+\frac{1}{6}\vec{l}_{5,n-1}
+\frac{1}{12}\vec{l}_{8,n-1}
+\frac{1}{24}L_{3,n-1}\vec{l}_{8,n-1}
\right),
\end{aligned}\displaybreak[0]\\
&\begin{aligned}
&S_{2,n,\text{Re}}
=\vec{\chi}_n^\dagger\Bigg(
\frac{2}{3}\vec{l}_{0,n-1}
+\frac{1}{6}\vec{l}_{3,n-1}
+\frac{1}{6}\vec{l}_{4,n-1}
+\frac{1}{6}\vec{l}_{5,n-1}
+\frac{1}{12}\vec{l}_{9,n-1}
+\frac{1}{12}\vec{l}_{7,n-1}\\
&+\frac{1}{6}L_{3,n-1}\bigg(
\vec{l}_{0,n-1}
+\frac{1}{2}\vec{l}_{4,n-1}
+\frac{1}{2}\vec{l}_{5,n-1}
+\frac{1}{4}\vec{l}_{9,n-1}
+\frac{1}{4}\vec{l}_{7,n-1}
\bigg)
\Bigg),
\end{aligned}\displaybreak[0]\\
&\begin{aligned}
&S_{3,n,\text{Re}}
=\vec{\chi}_n^\dagger\left(
\frac{1}{6}\vec{l}_{0,n-1}
+\frac{1}{6}\vec{l}_{4,n-1}
+\frac{1}{12}T_\text{Re}L_{2,n-1}\left(
\vec{l}_{2,n-1}
+\frac{1}{2}\vec{l}_{6,n-1}
\right)
\right),
\end{aligned}\displaybreak[0]\\
&\begin{aligned}
&S_{1,n,\text{Im}}
=\vec{\chi}_n^\dagger\left(
\frac{1}{6}\vec{m}_{0,n-1}
+\frac{1}{6}\vec{m}_{5,n-1}
+\frac{1}{12}\vec{m}_{8,n-1}
+\frac{1}{24}L_{3,n-1}\vec{m}_{8,n-1}
\right),
\end{aligned}\displaybreak[0]\\
&\begin{aligned}
&S_{2,n,\text{Im}}
=\vec{\chi}_n^\dagger\Bigg(
\frac{2}{3}\vec{m}_{0,n-1}
+\frac{1}{6}\vec{m}_{3,n-1}
+\frac{1}{6}\vec{m}_{4,n-1}
+\frac{1}{6}\vec{m}_{5,n-1}
+\frac{1}{12}\vec{m}_{9,n-1}
+\frac{1}{12}\vec{m}_{7,n-1}\\
&+\frac{1}{6}L_{3,n-1}\bigg(
\vec{m}_{0,n-1}
+\frac{1}{2}\vec{m}_{4,n-1}
+\frac{1}{2}\vec{m}_{5,n-1}
+\frac{1}{4}\vec{m}_{9,n-1}
+\frac{1}{4}\vec{m}_{7,n-1}
\bigg)
\Bigg),
\end{aligned}\displaybreak[0]\\
&\begin{aligned}
S_{3,n,\text{Im}}
=\vec{\chi}_n^\dagger\left(
\frac{1}{6}\vec{m}_{0,n-1}
+\frac{1}{6}\vec{m}_{4,n-1}
+\frac{1}{12}T_\text{Re}L_{2,n-1}\left(
\vec{m}_{2,n-1}
+\frac{1}{2}\vec{m}_{6,n-1}
\right)
\right).
\end{aligned}
\end{align}
\label{scalars_S_def}
\end{subequations}
\end{widetext}
\pagebreak
The above definitions allow us to write
\begin{subequations}
\begin{align}
&\begin{aligned}
&\vec{\chi}_n^\dagger\dpd{K_{n-1}}{a_p}\vec{\rho}_{n-1}
=\dpd{\text{Re}[\Omega]}{a_p}(t_{n-1})S_{1,n,\text{Re}}\\
&+\dpd{\text{Re}[\Omega]}{a_p}(t_{n-1}+(\Delta t)/2)S_{2,n,\text{Re}}\\
&+\dpd{\text{Re}[\Omega]}{a_p}(t_{n})S_{3,n,\text{Re}},
\end{aligned}\displaybreak[0]\\
&\begin{aligned}
&\vec{\chi}_n^\dagger\dpd{K_{n-1}}{b_p}\vec{\rho}_{n-1}
=\dpd{\text{Im}[\Omega]}{b_p}(t_{n-1})S_{1,n,\text{Im}}\\
&+\dpd{\text{Im}[\Omega]}{b_p}(t_{n-1}+(\Delta t)/2)S_{2,n,\text{Im}}\\
&+\dpd{\text{Im}[\Omega]}{b_p}(t_{n})S_{3,n,\text{Im}}.
\end{aligned}
\end{align}
\label{grad_F_def_single_term}
\end{subequations}

The backward propagation in Eq.~\eqref{adjoint_equation} can also be written 
explicitly. Define
\begin{subequations}
\begin{align}
&\vec{\mu}_{1,n} = L_{2,n-1}^\dagger\vec{\chi}_n,
&&\vec{\mu}_{2,n} = L_{3,n-1}^\dagger\vec{\chi}_n,\displaybreak[0]\\
&\vec{\mu}_{3,n} = L_{2,n-1}^\dagger\vec{\mu}_{1,n},
&&\vec{\mu}_{4,n} = L_{2,n-1}^\dagger\vec{\mu}_{2,n},\displaybreak[0]\\
&\vec{\mu}_{5,n} = L_{2,n-1}^\dagger\vec{\mu}_{4,n}.
\end{align}
\end{subequations}
Then
\begin{gather}
\begin{aligned}
&\vec{\chi}_{n-1}=\vec{\chi}_{n}\\
&+L_{1,n-1}^\dagger\left(
\frac{1}{6}\vec{\chi}_{n}
+\frac{1}{6}\vec{\mu}_{1,n}
+\frac{1}{12}\vec{\mu}_{3,n}
+\frac{1}{24}\vec{\mu}_{5,n}
\right)\\
&+\frac{2}{3}\vec{\mu}_{1,n}
+\frac{1}{6}\vec{\mu}_{2,n}
+\frac{1}{6}\vec{\mu}_{3,n}
+\frac{1}{6}\vec{\mu}_{4,n}
+\frac{1}{12}\vec{\mu}_{5,n}.
\end{aligned}
\label{adjoint_equation_expanded}
\end{gather}

\bibliographystyle{apsrev4-2-custom}
\bibliography{references}
\end{document}